\documentclass[11pt]{article}

\usepackage[utf8]{inputenc}
\usepackage{a4wide}
\usepackage[T1]{fontenc}
\usepackage[english]{babel}
\usepackage{enumerate}
\usepackage{xspace}
\usepackage{float}
\usepackage{adjustbox}
\usepackage{amsmath}
\usepackage{graphicx}
\usepackage{caption}
\usepackage{subcaption}
\usepackage[colorinlistoftodos]{todonotes}
\usepackage{url}
\usepackage[unicode=true,psdextra]{hyperref}
\usepackage{array}
\usepackage{tabularx}
\usepackage[titletoc,title]{appendix}
\usepackage{placeins}
\usepackage{comment}
\usepackage{adjustbox}
\usepackage{setspace}

\usepackage[T1]{fontenc}
\usetikzlibrary{positioning}
\usetikzlibrary{arrows, decorations.markings, calc}
\usepackage{subcaption}
\usetikzlibrary{arrows}
\usetikzlibrary{arrows.meta}

\makeatletter

\newif\ifinappendix
\inappendixfalse

\let\origappendix\appendix
\renewcommand{\appendix}{%
  \origappendix
  \inappendixtrue
}

\let\orig@seccntformat\@seccntformat
\renewcommand{\@seccntformat}[1]{%
  \ifinappendix
    \ifnum\pdfstrcmp{#1}{section}=0
      Appendix~\thesection.\quad
    \else
      \csname the#1\endcsname\quad
    \fi
  \else
    \orig@seccntformat{#1}%
  \fi
}

\makeatother

\usepackage[square,numbers]{natbib}
\hypersetup{
    colorlinks=true,
    linkcolor=red,
    filecolor=magenta, 
    citebordercolor=blue,
    citecolor=blue,     
    urlcolor=cyan,
    pdfauthor=Name
}

\usepackage{listings}
\usepackage{color}
\definecolor{dkgreen}{rgb}{0,0.6,0}
\definecolor{gray}{rgb}{0.5,0.5,0.5}
\definecolor{mauve}{rgb}{0.58,0,0.82}
\renewcommand{\thesection}{\arabic{section}}

\usepackage{listings}
\usepackage[a4paper]{geometry} 
\usepackage{mathtools,amssymb,bm,amsthm} 
\usepackage{xcolor} 
\usepackage{graphicx} 
  \graphicspath{{images/}} 
\usepackage{float} 
\usepackage{amsmath} 
\usepackage{caption} 

\usepackage{enumitem} 
\usepackage{pstricks-add}
\usepackage{pict2e} 
\usepackage{array}
\usepackage{multirow}
\usepackage{multicol}
\usepackage{ stmaryrd }
\usepackage{ marvosym }
\usepackage{pifont}
\usepackage{ upgreek }
\usepackage{comment}
\usepackage{ mathrsfs }
\usepackage{systeme}
\usepackage{ dsfont }
\usepackage{abstract}
\usepackage{indentfirst}

\usepackage{titlesec}

\titleformat{\paragraph}
{\normalfont\normalsize\bfseries}{\theparagraph}{1em}{}
\titlespacing*{\paragraph}{0pt}{3.25ex plus 1ex minus .2ex}{1.5ex plus .2ex}

\newtheorem{theorem}{Theorem}
\newtheorem{proposition}[theorem]{Proposition}

\newtheorem*{remark}{Remark}

\theoremstyle{definition}

\newcommand*\zgris{\mathbf{z}}
\newcommand*\Exp{\mathbb{E}}
\newcommand*\Var{\mathbb{V}}

\newcommand*\C{\textbf{C}}

\newcommand*{\cov}{\mathrm{cov}}
\newcommand*\Bell{\ensuremath{\boldsymbol\ell}}

\DeclareMathOperator*{\argmax}{argmax}
\DeclareMathOperator*{\argmin}{argmin}
\newcommand\R{\mathbb{R}}

\newcommand*\lik{\mathcal{L}}

\newcommand{\picut}{\pi_{\mathrm{cut}}}
\newcommand{\pifull}{\pi_{\mathrm{full}}}
\author{
  Oumar Baldé\footnotemark[1]\thanks{Université Paris-Saclay, CEA DES, Service de Génie Logiciel pour la Simulation, 91190, Gif-sur-Yvette, France}\,
  \footnotemark[2]\thanks{Institut de Mathématiques de Toulouse, Université Paul Sabatier, 118 Route de Narbonne, 31062 Toulouse. Oumar Baldé is currently a post-doctoral researcher at CEA, DES, IRESNE, DER, SESI.} 
  \and 
  Guillaume Damblin\footnotemark[1]\, \footnotemark[3]\thanks{Corresponding author.  (guillaume.damblin@cea.fr).}  
  \and 
  Amandine Marrel\footnotemark[4]\thanks{CEA DES, IRESNE, DER, SESI, Cadarache, 13108 St-Paul-Lez-Durance, France}\,
  \footnotemark[5]\thanks{LMA, Université d'Avignon, 84140 Avignon, France}
  \and  Antoine Bouloré \footnotemark[6]\thanks{CEA DES, IRESNE, DEC, SESC , Cadarache, 13108 St-Paul-Lez-Durance, France}
  \and Loïc Giraldi\footnotemark[6]
}

\title{A Gaussian process and linear-based method for computing cut distributions in modular Bayesian calibration of two chained computer models}

\date{\today}
\begin{document}
\maketitle
\begin{abstract}
Computer models are widely used in science and engineering to simulate complex systems. However, these models are affected by several sources of uncertainty, which may limit their use for decision making in risk management. We present a Bayesian approach for quantifying parameter uncertainty in a chain of two computer models motivated by multiphysics simulations in the nuclear field. Part of the inputs of a downstream model parametrized by $\theta \in \mathbb{R}^p$ come from the outputs of an upstream model parametrized by $\lambda \in \mathbb{R}^q$. Usually, the joint posterior distribution of $(\theta, \lambda)$ would be obtained by applying Bayes’ theorem using the experimental observations of both models. However, when the observations of the downstream model are too indirect to provide informative inference on $\lambda$, it may be preferable to compute a modular posterior distribution of $(\theta, \lambda)$, referred to as the \emph{cut distribution}. Assuming that the posterior distribution of $\lambda$ has been previously estimated from observations of the upstream model only, we aim to compute the posterior distribution of $\theta$ conditional on $\lambda$ using observations from the downstream model. To this end, we propose a Gaussian-process and linear-based framework to estimate the functional dependence between $\theta$ and $\lambda$, denoted by $\theta(\lambda)$, where each component is modeled as a realization of a Gaussian process. As the downstream model is approximated by a linear function of $\theta(\lambda)$, Bayesian conjugacy allows us to derive a Gaussian posterior predictive distribution of $\theta(\lambda)$ for any realization of $\lambda$. The effectiveness of the method is illustrated through several synthetic examples, and we highlight how variations in $\lambda$ impact the predictive distribution of the chained simulation.
\end{abstract}


\noindent\textbf{Keywords.} Bayesian calibration, cut distribution, chained models, Gaussian process.

\noindent\textbf{AMS classification:} 60G15, 62F15, 62G08.

\section{Introduction}
\label{sec:intro}

Numerical simulations have become essential for understanding, analyzing, and predicting complex systems and phenomena in all areas of engineering and science \citep{santner2018physical}. Indeed, when real field experiments are too costly or impossible to conduct for technical or ethical reasons, they are replaced by numerical counterparts that have benefited from a huge increase in computational resources over the last thirty years. However, the physical models and equations underlying the simulations are affected by various sources of uncertainty \citep{kennedy2001bayesian} that can affect the robustness of numerical predictions for decision-making. One of the most critical uncertainties is known as parameter uncertainty, which arises when a numerical simulation depends on a number of uncertain tuning or calibration parameters \citep{han2009simultaneous}. One popular way of inferring such parameters is Bayesian calibration using available experimental observations \citep{Sung24}. Bayesian calibration computes probability distributions for the uncertain parameters, unlike deterministic calibration, which provides a single best-fitting value \citep{Cam06}. 

\begin{figure}[ht!]
  \centering
  \label{fig:a}\includegraphics[width=\textwidth]{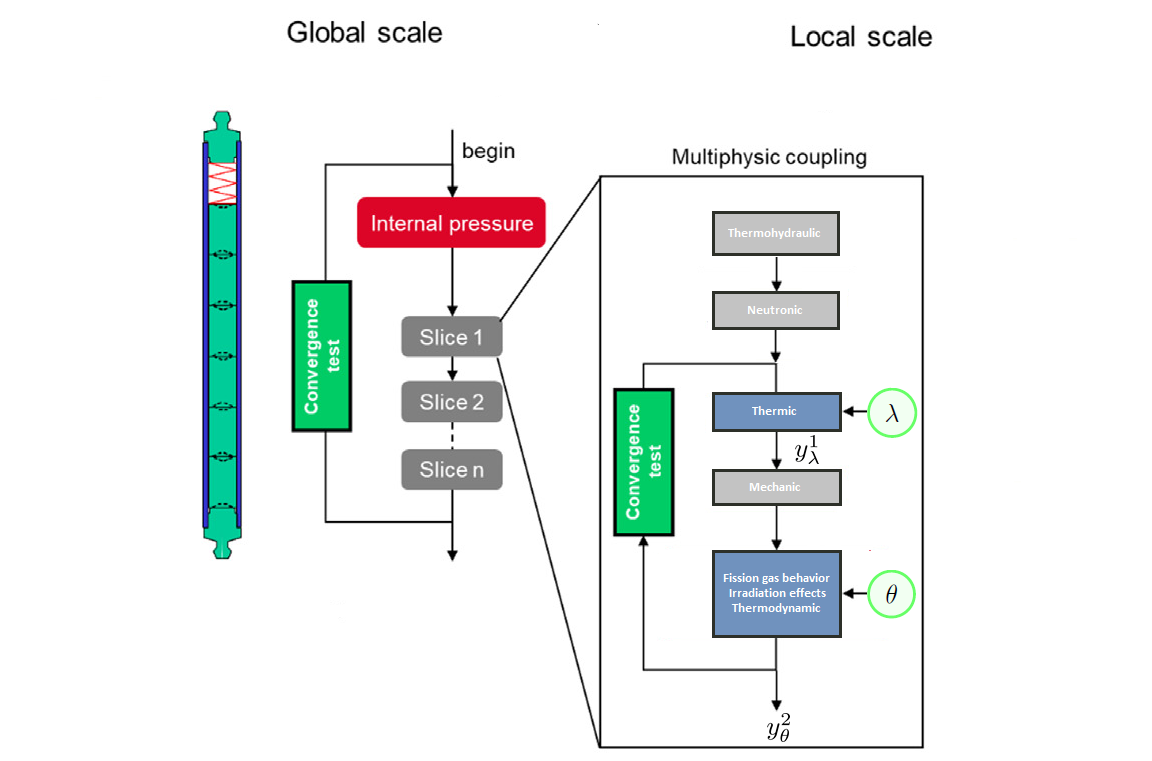}
  \caption{Chaining of the thermal and fission gas behavior models of the ALCYONE solver \citep{INTROINI2024110711}: $\lambda$ is the thermal conductivity and $\theta$  represents the parameters of the fission gas behavior model.}
  \label{fig1}
\end{figure}

This paper contributes to Bayesian calibration in the particular context of multiphysics simulations where several computer models of different physics are connected to one another to simulate the entire phenomenon of interest. As part of fuel simulations for nuclear power plants, we are interested in a multiphysics solver named ALCYONE \citep{INTROINI2024110711}, which is composed of interlinked models that represent the mechanical, thermal, and chemical behaviors of fuel rods in the core of pressurized water reactors. Recently, some papers have addressed forward uncertainty propagation for such multiphysics simulations in the nuclear field \cite{Delipei18, Zeng19, Avramova21}. Other works have dealt specifically with the emulation of a chain of several computer models via different strategies relying on Gaussian processes \cite{Kyzyurova18, Sanson19, Ming21}. For our part, we present a methodological contribution to Bayesian calibration for a chain of two models integrated into the ALCYONE solver. Specifically, we are interested in the fission gas behavior model, which takes as input the output of the thermal model. The latter simulates the evolution of the temperature within the fuel rod during the fission reaction and provides as output the associated temperature field. Then, the fission gas behavior model, as a function of the thermal model, continuously represents the behavior of the fission products (fuel swelling and release of fission gas atoms) during the fission reaction. Figure \ref{fig1} displays the two models in blue within the global multiphysics calculation workflow: the thermal model depends on the conductivity parameter $\lambda\in\Lambda\subset\R^{q}$ (here $q=1$) and the fission gas behavior model depends on the parameter $\theta\in\mathcal{T}\subset\R^{p}$ ($p\geq1$).

In the literature, several approaches have been proposed for quantifying parametric uncertainties in such a chain of numerical models.
First of all, there is the full modeling approach, which naturally conducts a simultaneous calibration of the two sets of parameters using all the available experimental
data related to both models \cite{DeCarlo16,marque2017calibration}. It has the advantage of dealing with all the uncertainties together, coherently, and using all available information. However, as pointed out and illustrated in \citep{jacob2017better}, the combination of several sources of information might lead to a misleading quantification of uncertainties. This happens, for example, when the observations of the downstream model are considered to be too indirect to bring valuable information for the posterior uncertainty of the parameters of the upstream model. An alternative is then to consider a two-stage calibration approach such as the segmented calibration \citep{DeCarlo16,Ye22} or the cut distribution \citep{plummer2015cuts,jacob2017better}. In these approaches, the parameters $\theta$ of the downstream model and the parameter $\lambda$ of the upstream model are, in a certain sense, calibrated separately with experimental observations representative of each model. Being unable to capture any statistical dependence between $\theta$ and $\lambda$, the segmented calibration may, however, inflate the predictive uncertainty of the downstream model. The cut distribution does not suffer from this limitation and, moreover, comes with statistical support that it may outperform the full posterior in certain settings. This typically occurs when some data bring information suspected to be not reliable enough because of model misspecification or lack of identifiability. Using such data may actually lead to lower predictive scores than those computed with a cut posterior \citep{jacob2017better}. Note that the latter distribution falls within modular Bayesian calibration approaches, as discussed originally in \citet{bayarri2009modularization}.

In this work, we aim to compute the cut distribution for quantifying the joint posterior uncertainty of $\theta$ and $\lambda$. On top of the previous discussion, choosing this framework is driven by the fuel application where a marginal posterior distribution of $\lambda$ was computed from an earlier calibration study based on the observations of the thermal model only. More importantly, this distribution has been validated by the fuel experts who developed the ALCYONE solver.
Therefore, we focus on the conditional distribution of $\theta$ given $\lambda$, denoted $\mathbb{P}_{\theta|\lambda}$, assuming that $\mathbb{P}_{\lambda}$ is known.
A naive approach to compute the cut distribution would then be to generate samples from $\mathbb{P}_{\theta|\lambda}$ for a large set of samples $\lambda$ drawn from $\mathbb{P}_{\lambda}$. Unfortunately, this sampling scheme is computationally expensive and does not leverage the fact that, under some regularity conditions, the conditional distribution of $\theta$ given $\lambda$ may provide some information about $\theta$ given $\lambda^{\prime}$ if $\lambda$ and $\lambda^{\prime}$ are close to each other. Instead, we present another approach that directly computes the probability distribution of the functional parameter $\theta(\lambda)$. This nonparametric approach represents each component of $\theta(\lambda)$ as a trajectory of a Gaussian process, which is inspired by the work of \citep{brown2018nonparametric} in another calibration context. In the present paper, the functional approach will be presented in the case where the output of the downstream model is expressed as a linear function of $\theta$. This assumption makes it feasible to derive analytically the posterior predictive distribution of $\theta$ conditional on any realization of $\lambda$ drawn from $\mathbb{P}_{\lambda}$. The performance of the approach will be illustrated numerically on several academic examples.

The paper is organized into five parts presenting the statistical framework of the functional approach, while its implementation on the two ALCYONE solver models is deferred to future work. Section \ref{sec:bayesian} introduces Bayesian calibration of two chained numerical models, including a brief review of the seminal work of Kennedy and O'Hagan and a presentation of the cut posterior distribution. Section \ref{sec:approach} deals with several possible methods for conditional density estimation. In Section \ref{sec:gplincc}, the functional approach called GP--LinCC for Gaussian process and linear-based Conditional Calibration is developed. Its numerical performance will be demonstrated in Section \ref{sec:num}. Section \ref{sec:conclusion} ends the paper with some conclusions and perspectives.

\section{Bayesian calibration of two chained numerical models}
\label{sec:bayesian}
\subsection{The full posterior versus the cut posterior}

We deal with a computer model $y$ that is supposed to be representative of a physical system of interest $r$. The latter yields a scalar quantity of interest $r(x)$, where $x \in \mathcal{X} \subset \mathbb{R}^{d}$ denotes an input vector composed of control variables such as boundary and initial conditions. Let us assume that the computer model is written as the composition of two submodels. Then
\begin{equation}
\label{chain}
y_{\theta, \lambda}(x):=y^2_\theta(y^1_\lambda(x))
\end{equation}
with the input variables of the downstream model being the outputs of the upstream model. The computer model often needs to be parameterized by calibration or tuning parameters, which may have no direct experimental counterpart \cite{loeppky2006computer,han2009simultaneous}. In Eq.~(\ref{chain}), $\theta$ and $\lambda$ refer to the parameters of the downstream and upstream models respectively. Being uncertain, these parameters are usually estimated to obtain the best agreement between the computer model and the physical system. This procedure is known as model calibration and relies on the availability of experimental observations of $r(x)$. For $1\leq i\leq n$, an experimental observation $z_i$ at a specific input location $x_i$ is in fact related to $r(x_i)$ by the equation
\begin{equation}
\label{exp_vs_reality}
z_i=r(x_i)+\epsilon_{z_i}
\end{equation}
where $\epsilon_{z_i}$ is the realization of a zero-mean Gaussian distribution representing the experimental uncertainty. Then, the computer model replaces $r(x)$ in Eq.~(\ref{exp_vs_reality}), up to a discrepancy function $b$ \citep{kennedy2001bayesian}:
\begin{equation}
    z_i= y_{\theta, \lambda}(x_i) + b(x_i) +\epsilon_{z_i}, \quad 1 \le i \le n,
    \label{eq1}
\end{equation}
The function $b(x)$, called model discrepancy, is originally presented by Kennedy and O'Hagan in \citep{kennedy2001bayesian} to represent the gap between the numerical model $y_{\theta, \lambda}(x)$ and the physical system $r(x)$ when the model is run at the optimal (but unknown) value $(\theta, \lambda)$ of the parameters\footnote{Optimal in the sense that the model run with $(\theta, \lambda)$ yields the best possible predictive accuracy. Note that this optimal value may differ from the true parameter \citep{kennedy2001bayesian, wu2018inverse}.}
. If $b(x)$ is judged negligible compared to the experimental uncertainty \citep{damblin2018adaptive}, the following simplified equation can be chosen instead$:$
\begin{equation}
    z_i = y_{\theta, \lambda}(x_i) +\epsilon_{z_i}.
    \label{eq2}
\end{equation}
Assuming that the standard deviation of $\epsilon_{z_i}$, denoted $\sigma_{\epsilon_{z_i}}$, is known, the joint posterior distribution of $(\theta,\lambda)$ based on Eq.~(\ref{eq2}) is obtained via Bayes' theorem as
\begin{equation}
     \pi(\theta,\lambda|z)\propto \lik(z|\theta, \lambda)\,\pi(\theta, \lambda),
    \label{eq3}
\end{equation}
where $\pi(\theta, \lambda)$ is the prior density that quantifies the uncertainty of $(\theta,\lambda)$ before collecting the data $z$, $\lik(z|\theta,\lambda)$ is the likelihood of the data $z$ conditional on the pair $(\theta,\lambda)$ and $\pi(\theta,\lambda| z)$ is the posterior density that quantifies the residual uncertainty of $(\theta,\lambda)$ conditional on $z$. Note that the same symbol $\pi$ is used on both sides to denote, respectively, the prior and posterior densities. This notation avoids introducing $\pi_{\text{prior}}(\cdot)$ and $\pi_{\text{post}}(\cdot)$ and does not imply that the two densities correspond to the same function.

Eq.~(\ref{eq2}) relies on experimental data from the last stage of the simulation chain. 
When additional data informing the parameter $\lambda$ of the thermal model are available, cut-off models can be used. 
They explicitly partition the different sources of information contributing to the identification of $(\theta,\lambda)$ \citep{bayarri2009modularization, plummer2015cuts, jacob2017better}. Inspired by the work of \citet{plummer2015cuts}, Figure~\ref{fig2} presents a cut-off model for the chaining in Eq.~(\ref{chain}) where the direct measurements $w$ (a realization of some random variable $W$) bring information about $\lambda$ through comparisons with the outputs of the upstream model$:$
\begin{equation}
w_j = y^1_{\lambda}(x_j) + \epsilon_{w_j}, \quad 1 \le j \le n_1,
\end{equation}
where $\epsilon_{w_j}$ is still the realization of a zero-mean Gaussian distribution. In this figure, the graph is partitioned by a cut between the two models, preventing the data $z$ (a realization of $Z$) from influencing the estimation of $\lambda$.
On the left-hand side, the posterior distribution of $\lambda$ is computed independently of the data $z$. In other words, the resulting estimate relies exclusively on the observation $w$ of $W$, despite the additional information provided by $Z$.

\begin{figure}[ht!]
\centering
\begin{tikzpicture}[
  >=latex,
  thick,
  node distance=1.6cm and 2.2cm,
  font=\sffamily,
  round/.style={
    circle,
    draw=green!50!black,
    fill=green!5,
    very thick,
    minimum size=10mm,
    align=center
  },
  roundbig/.style={
    circle,
    draw=green!50!black,
    fill=green!5,
    very thick,
    minimum size=22mm,  
    align=center
  },
  square/.style={
    rectangle,
    draw=green!50!black,
    fill=green!5,
    very thick,
    minimum height=9mm,
    minimum width=14mm,
    align=center
  }
]

\node[roundbig] (M1)                      {Model 1\\[-0mm] $y^1_{\lambda}(x)$};
\node[round]    (lbd)   [above=of M1]     {$\lambda$};
\node[square]   (W)     [below=of M1]     {$W$};

\node[roundbig] (M2)    [right=3.2cm of M1] {Model 2\\[-0mm] $y^2_{\theta}(y^1_{\lambda}(x))$};
\node[round]    (theta) [above=of M2]     {$\theta$};
\node[square]   (Z)     [below=of M2]     {$Z$};

\node[round]    (epsW)  [left=of W]       {$\epsilon_W$};
\node[round]    (epsZ)  [right=of Z]      {$\epsilon_Z$};

\draw[-Implies, double distance=1pt] (lbd) -- (M1);
\draw[->]                            (M1) -- (W);
\draw[-Implies, double distance=1pt] (theta) -- (M2);
\draw[->]                            (M2) -- (Z);

\draw[->] (epsW) -- (W);
\draw[->] (epsZ) -- (Z);

\draw[-Implies, double distance=1pt] 
    (M1) -- node[midway, fill=white, inner sep=1.5pt]
    {\textcolor{red}{\Large$\pmb{\times}$}} (M2);

\end{tikzpicture}

\caption{Graphical representation of a cut-off model for two chained models. 
Single and double arrows, respectively, denote stochastic and deterministic functional dependencies.
Rectangular nodes represent observed data; the red cross indicates that information from $Z$ 
does not back-propagate to $\lambda$.}
\label{fig2}
\end{figure}
\noindent Then, we can write the probability distribution of the parameters $(\theta, \lambda)$ conditional on the complete data $(W=w, Z=z)$ as in Eq.~(\ref{eq3}) of \cite{jacob2017better}:
\begin{equation}
     \picut(\theta,\lambda | w,z)=\pi(\theta | \lambda,z)\,\pi(\lambda | w),
     \label{eq4}
\end{equation}
where
\begin{equation}
\pi(\lambda| w) \propto   \lik(w | \lambda) \pi(\lambda)
\end{equation}
is the posterior distribution of $\lambda$ with respect only to the data $w$ of the upstream model, and $\pi(\theta | \lambda,z)$ is the posterior distribution of $\theta$ conditional on $\lambda$ with respect to the data $z$ of the downstream model. The density $\pi_{\mathrm{cut}}(\theta,\lambda | w,z)$, called the cut distribution in \citep{plummer2015cuts}, does not coincide with the regular joint posterior density, denoted by $\pi_{\mathrm{full}}(\theta,\lambda | w,z)$, which is written as:
\begin{equation}
\begin{array}{ccc}
       \pifull(\theta,\lambda | w,z)&=&\pi(\theta | \lambda,z)\,\pi(\lambda | w,z).
\end{array}
    \label{eq5}
\end{equation}
It turns out that these two distributions are linked by the following equations:
\begin{equation}
	\frac{\pifull(\theta,\lambda | w,z)}{\picut(\theta,\lambda | w,z)}
 = \frac{\pi(\lambda | w,z)}{\pi(\lambda | w)} = \frac{\pi(z | \lambda)}{\pi(z | w)}.
	   \label{eq6}
\end{equation}

\subsection{Illustrative numerical example for a simple calibration problem}

\label{cut_vs_full_simple_example}    
Let us consider the simple analytical example below, inspired by DeCarlo et al.~\cite{DeCarlo16}. 
For $1\leq j\leq n_1$ and $1\leq i\leq n=n_2$,
\begin{equation}
\label{toy_example}
\left\{
\begin{array}{lll}
\displaystyle  w_j = \lambda +\epsilon_{w_j} & ; & \epsilon_{w_j} \sim \mathcal{N}(0, \sigma^2_{w}), \\
\displaystyle z_i = x_i\lambda + \theta + \epsilon_{z_i} & ; & \epsilon_{z_i} \sim \mathcal{N}(0, \sigma^2_{z}),
\end{array}
\right.
\end{equation}
with $\sigma^2_{z} = 0.15$ and $\sigma^2_{w} = 0.15$.
Eq.~(\ref{toy_example}) can be rewritten as
\begin{equation}
y = 
\begin{pmatrix}
w \\
z
\end{pmatrix}
= 
A_x
\begin{pmatrix}
\lambda \\
\theta
\end{pmatrix}
+
\begin{pmatrix}
\epsilon_w \\
\epsilon_z
\end{pmatrix},
\end{equation}

with
\begin{equation}
\begin{array}{ccc}
A_{x} = 
\begin{pmatrix} 
\mathbf{1}_{n_1} & \mathbf{0}_{n_1} \\ 
x & \mathbf{1}_{n_2} 
\end{pmatrix}
& ; & 
x =(x_1, \cdots, x_{n_2})^t
\end{array}
\end{equation}

and
\begin{equation}
\begin{pmatrix} \epsilon_w \\ \epsilon_z \end{pmatrix} 
\sim 
\mathcal{N}_{n_1+n_2}\!\left(0, 
\Sigma_{\sigma}:=
\begin{pmatrix} 
\sigma^2_w I_{n_1}  & 0 \\ 
0 & \sigma^2_z I_{n_2} 
\end{pmatrix}
\right).
\end{equation}

The impact of two different prior distributions on $\pi_{\mathrm{full}}$ and $\pi_{\mathrm{cut}}$ will be assessed:
\begin{enumerate}
\item An informative Gaussian prior on $\xi := (\lambda,\theta)^{\top}$ given by
\begin{equation}
\xi \sim
 \mathcal{N}\left(
 \xi_0 := \begin{pmatrix} \lambda_0 \\ \theta_0 \end{pmatrix},
 \Sigma_0 :=
 \begin{pmatrix} \sigma^2_{\lambda_0} & 0 \\ 0 & \sigma^2_{\theta_0} \end{pmatrix}
 \right).
 \label{eqGaussianPrior}
\end{equation}

\item A Jeffreys prior on $\xi$, i.e.,
\begin{equation}
\pi(\xi) \propto 1.
 \label{eqJeffreysPrior}
\end{equation}

\end{enumerate}

\begin{figure}[ht!]
    \centering
    \includegraphics[width=0.7\textwidth]{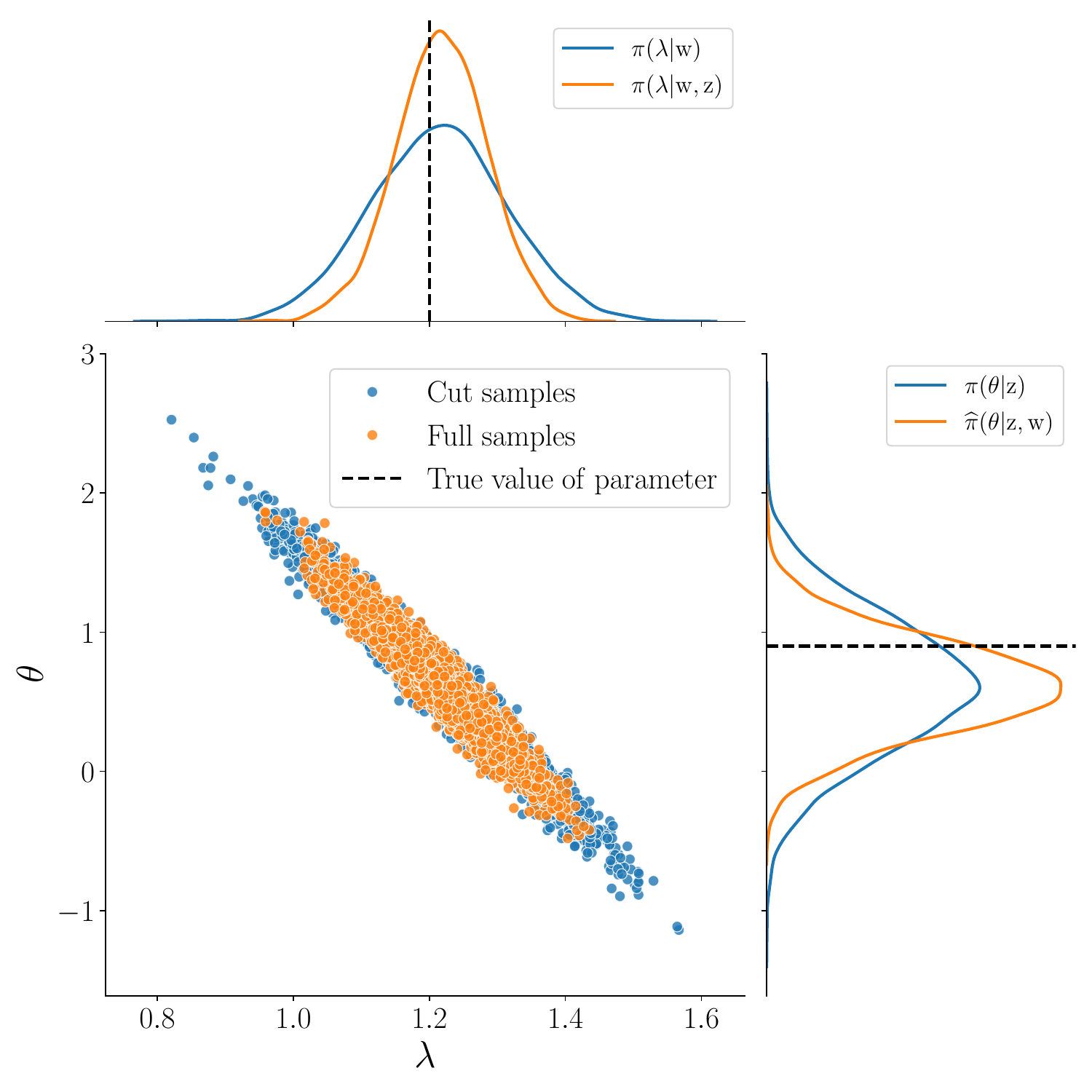}
\caption{
Comparison of the joint distribution of $(\theta,\lambda)$ obtained from the cut posterior $\picut(\theta,\lambda | w,z)$ (red) and the full posterior $\pifull(\theta,\lambda | w,z)$ (blue) under the Gaussian prior $\pi(\theta,\lambda)$ of Eq. \eqref{eqGaussianPrior} with $\lambda_0=1$, $\theta_0=0.7$ and $\sigma^2_{\lambda_0}=1$, $\sigma^2_{\theta_0}=0.2$. The notation $\widehat{\pi}(\cdot | v)$ refers to the kernel density estimator of $\pi(\cdot | v)$.
}
\label{figExample_model1}
\end{figure}

The expressions of both the full and cut posterior distributions can be explicitly derived, as well as the KL divergence between the two (see Appendix \ref{analytical}). The following results have been established:

\begin{enumerate}
\item With Gaussian prior:
\begin{equation}
\label{Gaussian_prior_KL}
\mathrm{KL}\!\left(\pi(\lambda | w,z) \parallel \pi(\lambda | w)\right)
= 0 
\iff 
\forall\, i,\; x_i = 0.
\end{equation}

\item With Jeffreys prior:
\begin{equation}
\label{Jeffreys_prior_KL}
\mathrm{KL}\!\left(\pi(\lambda | w,z) \parallel \pi(\lambda | w)\right)
= 0 
\iff 
\forall\, i,\; x_i = c \in \mathbb{R}.
\end{equation}
\end{enumerate}
Eq.~(\ref{Gaussian_prior_KL}) follows directly from the fact that the two models are no longer linked to one another. Nonidentifiability means that multiple pairs $(\theta,\lambda)$ can yield identical predictions for the downstream model, as occurs when all $x_i$ are equal to a constant $c$. In this setting, Eq.~(\ref{Jeffreys_prior_KL}) shows that nothing is gained by computing $\pi_{\mathrm{full}}$ instead of $\pi_{\mathrm{cut}}$ when a Jeffreys prior is used. When a Gaussian prior is used instead, Eq.~(\ref{Gaussian_prior_KL}) states that the KL divergence is nonzero. To illustrate this,
we generated the data $w$ and $z$ with sample sizes $n_1 = n_2 = 15$, the true parameter values
$\theta = 0.9$, $\lambda = 1.2$, $c = 5$, and variances $\sigma_z^2 = 0.15$ and $\sigma_w^2 = 0.15$. Figure~\ref{figExample_model1} displays the histograms of $\picut(\theta,\lambda | w,z)$ and
$\pifull(\theta,\lambda| w,z)$ together with a scatter plot of the 2-D posterior samples. Although the gap between the two distributions is moderate, $\pi_{\mathrm{full}}$ outperforms $\pi_{\mathrm{cut}}$ because the chosen prior density for $\theta$ agrees sufficiently well with the data $z$. However, this behavior is not systematic, and the full posterior distribution might be less accurate than the cut distribution, depending on the interaction between the value of $c$ and the shape of the prior, as shown in \cite{bayarri2009modularization, jacob2017better}. 

\begin{figure}[ht!]
    \centering
    \includegraphics[width=0.7\textwidth]{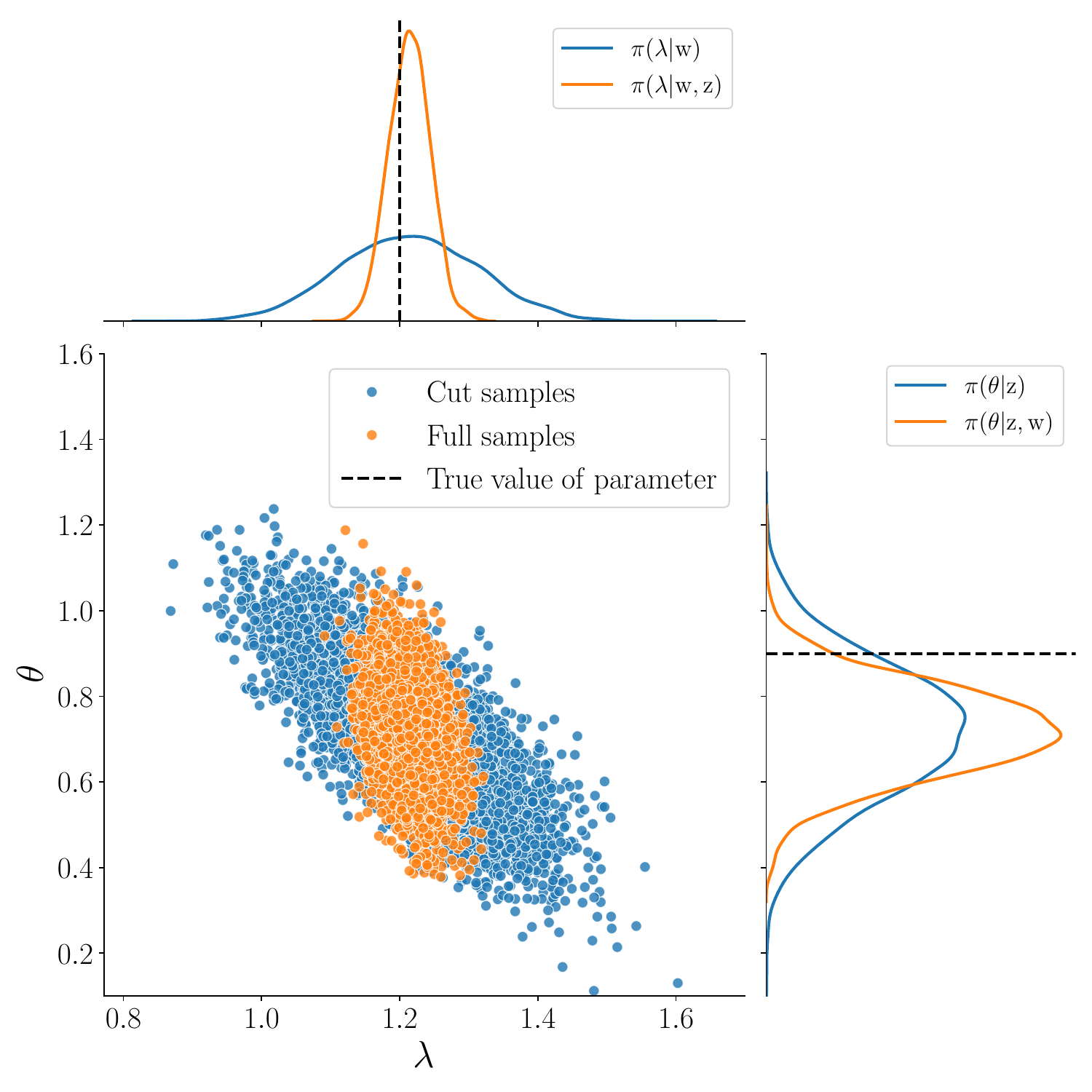}
\caption{
Comparison of the joint distribution of $(\theta,\lambda)$ obtained from the cut posterior 
$\picut(\theta,\lambda | w,z)$ (red) and from the full posterior 
$\pifull(\theta,\lambda | w,z)$ (blue) under a Jeffreys prior $\pi(\theta,\lambda)$ of Eq. \eqref{eqJeffreysPrior}.
}

    \label{figExample_model2}
\end{figure}

\medskip
We now turn to the identifiable setting, that is, when $x_i$ varies. A more significant gap between $\pi_{\mathrm{full}}$ and $\pi_{\mathrm{cut}}$ appears, regardless of the prior type. Figure \ref{figExample_model2} shows the posterior distributions when the Jeffreys prior is used and $z$ is simulated with
\begin{equation}
x_i \sim \mathcal{U}[-6, 6],\qquad 1\leq i\leq n_2=15
\end{equation}
and with the same values as before for $\theta$, $\lambda$, $\sigma_z^2$, and $\sigma_w^2$. Since the data $z$ now provide additional information about $\lambda$, using $\pi_{\mathrm{full}}$ will always yield a significantly more precise estimate of $\lambda$. We can also see that the posterior covariance between $\lambda$ and $\theta$ is strongly modified.


\medskip
In these well-specified synthetic examples, i.e., where the data-generating process coincides with the model used for inference, we have illustrated the impact of downstream model identifiability on the shape of $\pi_{\mathrm{full}}$ and $\pi_{\mathrm{cut}}$. 

Under model misspecification, even when the downstream model is identifiable, Section~3 of \cite{jacob2017better} shows that the full posterior may yield suboptimal predictive performance, whereas modular strategies such as the cut distribution can lead to better-calibrated predictions. This situation typically arises when a model-discrepancy term $b(x)$ in Eq.~(\ref{eq1}) exists between the outputs of the chained models and the corresponding observations but is omitted in the Bayesian calibration process. Introducing $b(x)$ would substantially increase inference complexity and may implicitly reveal structural limitations of the numerical model, which partly explains its limited adoption in engineering practice. In fuel-performance simulations where current calibration practices do not incorporate $b(x)$, the cut distribution is therefore particularly appropriate. A final argument in favor of the cut posterior is that the experimental data $z$ associated with the fission-gas behavior model provide much weaker information about $\lambda$ than the thermal-model data $w$. The contribution of $z$ may thus be counterproductive, making the full posterior less accurate than the cut posterior for estimating $\lambda$.

\medskip
In the rest of the paper, we develop a new method, called GP--LinCC (Gaussian process and linear-based Conditional Calibration), to compute the cut posterior distribution when the posterior density $\pi(\lambda|w)$ is known. This reduces to estimating the conditional posterior density $\pi(\theta|z,\lambda)$. The next section highlights the limitations of existing approaches for this task and introduces the key components of GP--LinCC.

\section{Methods for conditional density estimation}
\label{sec:approach}
\subsection{Nonparametric approach via a kernel density estimation (KDE)}

An estimate of the conditional density $\pi(\theta|\lambda,z)$ can be defined as:
\begin{equation}
\label{kde}
\widehat{\pi}(\theta|\lambda,z)=
\frac{\widehat{\pi}_{\mathrm{cut}}(\theta,\lambda|w,z)}{\pi(\lambda|w)}
\end{equation}
where $\widehat{\pi}_{\mathrm{cut}}$ denotes a KDE estimate \cite{Li2007,Otneim16} of 
$\pi_{\mathrm{cut}}(\theta,\lambda|w,z)$. While in standard conditional density estimation the denominator in Eq.~(\ref{kde}) is also approximated using KDE \cite{Chen01}, $\pi(\lambda|w)$ is known in our setting, which simplifies the ratio estimation.
The numerator is approximated from a large number of samples $\{\theta_i,\lambda_i\}_{i=1}^N$ drawn from $\pi_{\mathrm{cut}}(\theta,\lambda|w,z)$. 
These samples can be generated using a Gibbs sampler from the class of Markov chain Monte Carlo (MCMC) algorithms \citep{robert1999monte}. 
Starting from an initial value $\lambda_0$, the algorithm proceeds as follows:
for $1\leq i\leq N$,
\begin{align}
    & \theta_i\sim\pi(\theta| \lambda_{i-1},z)
      \propto \pi_{\text{full}}(\theta,\lambda_{i-1}| w,z), \\
    & \lambda_i\sim \pi(\lambda| \theta_i,z)
      \propto \frac{ \pi_{\text{full}}(\theta_i,\lambda| w,z) }
                   {\pi(z| \lambda)}.\label{eq8}
\end{align}
$N$ should be large enough to allow convergence of this sampling
scheme towards the cut distribution.
Unfortunately, in the general case, the marginal likelihood in the denominator
of Eq.~\eqref{eq8}, called the \textit{feedback} term in \cite{jacob2017better},
has no closed form:
\begin{equation}
    \pi(z| \lambda)=\int_{\mathcal{T}} \lik(z| \theta,\lambda)\,\pi(\theta| \lambda)\, d\theta.
\end{equation}
Unless a complex approximation of the feedback term is used \cite{Liu22},
the above sampling scheme is infeasible.
Instead, a simple approach implemented in \citet{Jacob20} may be to sample,
for any realization $\lambda_i$ drawn from $\pi(\lambda| w)$,
the associated conditional posterior distribution:
for $1\leq i\leq N$,
\begin{align}
    &\lambda_i \sim \pi(\lambda| w),\\
    &\theta_i \sim \pi(\theta| \lambda_i,z). \label{naive}
\end{align}
If the conditional density in Eq.~(\ref{naive}) is known only up to a constant,
drawing one sample $\theta_i$ requires a specific MCMC algorithm depending on $\lambda_i$.
This sampling scheme thus involves as many Markov chains as the number of realizations
$\lambda_i$ drawn from $\pi(\lambda| w)$, and the convergence of each Markov chain
must be diagnosed carefully to ensure accurate sampling of the cut distribution.
Moreover, if the computer model is even moderately time-consuming, the total number of generated samples $(\lambda_i,\theta_i)$ may be insufficient to construct a reliable KDE.
This motivates the moment-based method presented below, where a regression model fits
the conditional density $\pi(\theta| \lambda,z)$ as a function of $\lambda$.

\subsection{Moment-based estimation method}
\noindent Let us consider a numerical design, denoted by $D_m$, consisting of $m$ realizations $\lambda_j \sim \pi(\lambda|w)$ generated by Latin hypercube sampling \citep{mckay1979comparison}. Given a set of posterior samples $(\theta^{(i)}_j)_{i=1}^N$ 
associated with each $\lambda_j$, the posterior expectation and the variance matrix of the 
conditional distribution in Eq.~(\ref{naive}) can be estimated by
\begin{align}
    &\bar{\theta}(\lambda_j):=\Exp(\theta|\lambda_j,z)\approx \frac{1}{N}\sum_{i=1}^N\theta^{(i)}_j,\label{eq10}\\
    &\cov(\theta|\lambda_j,z)\approx \frac{1}{N}\sum_{i=1}^N (\theta^{(i)}_j- \bar{\theta}(\lambda_j) )
      (\theta^{(i)}_j- \bar{\theta}(\lambda_j) )^t.\label{eq11}
\end{align}
Then, a Gaussian process (GP) emulator (see Section~\ref{sec::gp} for a brief introduction) can be used to interpolate the first two moments \eqref{eq10} and \eqref{eq11}, enabling predictions of the mean and variance of $\mathbb{P}_{\theta|\lambda}$ for any realization $\lambda^{\star} \notin D_m$. The main problem lies in preserving the positive semidefinite property of the variance matrix. The GP can be fitted 
on the log variance if $\theta$ is a scalar parameter $(p=1)$. For $p>1$, the solution proposed in 
\citep{fiszeder2021covariance} consists of fitting a GP on each Cholesky factor, then using the 
inverse Cholesky decomposition to obtain a matrix ensuring the positive semi-definite property. 
However, this method is rather costly when $p$ is large and does not provide any uncertainty of the 
predicted matrix. 

\medskip
In the rest of the paper, the downstream model is approximated by a linear function of $\theta$ conditional on $\lambda$ (see Section~\ref{sec:linear_app}). This assumption is appropriate when the model output can be reasonably approximated by a linear function of $\theta$, as in the case of the fission-gas behavior model. This implies that Eq.~(\ref{naive}) no longer requires MCMC, although a large number of samples $(\lambda_i,\theta_i)$ is still needed. In the linear 
framework, the moment-based method becomes easier to implement because the posterior of $\theta$ 
conditional on $\lambda$ can be computed explicitly as a Gaussian distribution provided the prior 
density $\pi(\theta|\lambda)$ is Gaussian. However, the difficulty of interpolating variance matrices remains. In the GP--LinCC method presented in the next section, we adopt a Bayesian approach by fitting a GP emulator embedded as a prior distribution on $\theta(\lambda)$.

\subsection{Proposed solution: method based on GP-prior and linear assumption (GP--LinCC method)}

\subsubsection{Gaussian Process prior}
\label{sec::gp}

Gaussian processes (GPs) are widely used to emulate computationally expensive black-box computer models \citep{sacks1989design}. A GP defines a prior over the response function of such models, fully characterized by a mean function and a covariance kernel. 
Conditioned on observed data, the resulting posterior GP yields a Gaussian predictive distribution for the model output at any input location, with closed-form expressions for the predictive mean and covariance matrix \citep{currin1991bayesian} (see Appendix~\ref{gp}).
In nuclear computational modeling, GPs have been used to address inverse uncertainty quantification \citep{Damblin20}, to identify penalizing configurations for safety studies \citep{marrel2022icscream}, and for various other applications. In a context close to that of this paper, GPs were used to emulate a chain of two numerical models for calibration 
purposes \citep{sophie2016}. Still in the context of model calibration, a GP has been used in \citep{brown2018nonparametric} to capture the 
functional relationship between a calibration parameter and some input control variables. In our framework, inspired by the latter reference, we model the relationship between $\theta$ and $\lambda$ by assuming that each component of $\theta(\lambda)\in\mathbb{R}^p$ follows an independent GP a priori, such that
\begin{equation}
    \theta_u(\lambda)\overset{\mathrm{indep.}}{\sim} 
    \mathcal{GP}\big(m_{\beta_u}(\lambda),\, \sigma_u^2 K_{\psi_u}(\lambda,\,\lambda^\prime)\big),\,
    1 \leq u \leq p,
    \label{eq12}
\end{equation}
where $m_{\beta_u}(\lambda)$ is the mean function (also called trend) of the $u$th GP. A constant 
$m_{\beta_u}(\lambda)=\beta_u$ or a degree-one polynomial trend is commonly used in practice.

\medskip
For simplicity, we assume in the sequel that the prior mean is constant and equal to $\beta_u$. 
The covariance function $\sigma_u^2 K_{\psi_u}(\lambda,\lambda^\prime)$ controls both the regularity and the scale of the GP trajectories. 
It encodes the dependence structure of the $u$th GP and must be positive semidefinite. 
When $\theta_u(\lambda)$ is assumed to be highly smooth, the Matérn~5/2 covariance function is among the most commonly used choices, as recommended in particular in \citep{gu2018robust}. 
It is defined as
\begin{equation}
   \sigma_u^2 K_{\psi_u}(\lambda,\lambda^\prime)
   = \sigma_u^2 
     \left(1+\sqrt{5}\frac{|\lambda-\lambda^\prime|}{\psi_u}
     + \frac{5}{3}\left(\frac{|\lambda-\lambda^\prime|}{\psi_u}\right)^2\right)
     \exp\!\left(-\sqrt{5}\frac{|\lambda-\lambda^\prime|}{\psi_u}\right).
   \label{matern}
\end{equation}
In the multidimensional case (i.e., $\lambda \in \mathbb{R}^{q}$ with $q\geq 1$), either an isotropic or a separable (tensor-product) Matérn~5/2 covariance function may be employed (see, for example, \cite{Bachoc13}).

\subsubsection{Linear approximation}
\label{sec:linear_app}

The advantage of a linear framework is that the posterior distribution of $\theta$ conditional on 
$\lambda$ can be derived analytically. We assume that, for any $\lambda$, the output of the numerical 
model $y_{\theta(\lambda),\lambda}(x_i)$ can be written, or approximated, as a linear function of 
$\theta(\lambda)$. Thus, for $1 \leq i \leq n$, Eq.~\eqref{eq2} can be rewritten as
\begin{equation}
 z_i = g_{\lambda,0}(x_i) + g_{\lambda,1}(x_i)^t \theta(\lambda) + \epsilon_{i,\lambda},
 \qquad \epsilon_{i,\lambda} \sim \mathcal{N}(0,\,\sigma^2_{\epsilon_i} + \delta^2_{\lambda,i}),
 \label{eq13}
\end{equation}
where $\delta^2_{\lambda,i}$ is a scale parameter capturing the discrepancy between the model output 
and its linear approximation. In practice, the regression coefficients are collected in the vector
\begin{equation}
g_\lambda(x_i):=\big(g_{\lambda,0}(x_i),\, g_{\lambda,1}(x_i)\big)^t \in \mathbb{R}^{p+1}.
\end{equation}
These coefficients must be estimated at fixed $(\lambda, x_i)$, either by performing a local linearization 
at a specific $\theta(\lambda)=\tilde{\theta}(\lambda)$ or by using a variational linear approximation over 
$\mathcal{T} \ni \theta$. We adopt the latter approach, as advocated in a recent paper addressing the 
identifiability of inverse problem solutions \cite{Bousquet25}. A linear regression model is then fitted 
for each pair $(\lambda_j, x_i)$ using a set of $n_{\mathrm{sim}}$ training samples defined as
\begin{equation}
\Theta^{(j)}:=\{\theta^{(j)}_{1},\ldots,\theta^{(j)}_{n_{\mathrm{sim}}}\}\subset\mathcal{T}^{n_{\mathrm{sim}}}.
\end{equation}
A total of $m \times n \times n_{\mathrm{sim}}$ simulations are therefore required to fit the 
$m \times n$ linear models.

\section{The GP-LinCC method}
\label{sec:gplincc}
Building on the linear framework introduced previously, Section~\ref{sec:post} establishes that combining the Gaussian prior in Eq.~\eqref{eq12} with a Gaussian likelihood yields a Gaussian posterior distribution. Section~\ref{sec:pred} further shows that this structure leads to a Gaussian predictive distribution for $\theta$ conditional on any $\lambda$ with non-zero probability under $\pi(\lambda|w)$. 

The practical implementation of the GP--LinCC method needs the following set of simulations of the chain model:
\begin{equation}
\cup_{j=1}^{m}\cup_{k=1}^{n_{\mathrm{sim}}}\{y_{\theta^{(j)}_k,\,\lambda_j}(x_i)\}_{i=1,\ldots,n}.
\end{equation}
These simulations make it possible to estimate all the vectors $g_{\lambda_j}(x_i)$ for $1 \leq i \leq n$ and $1 \leq j \leq m$, and thus to fit the $m \times n$ linear models required by the approach.

\subsection{Posterior inference}
\label{sec:post}

We apply Eq.~\eqref{eq13} to the $m$ realizations $\lambda_j$ to learn the relation between 
$\theta$ and $\lambda$. We can then write $m$ equations involving the experimental data 
$z = (z_1,\ldots,z_n)^t \in \mathbb{R}^{n}$:
\begin{equation}
    z = g_{\lambda_j}(x)\,\theta(\lambda_j) + \epsilon_{\lambda_j}, 
    \qquad 1 \le j \le m,
    \label{eq15}
\end{equation}
where 
\[
g_{\lambda_j}(x)
:= \big(g_{\lambda_j}(x_1)^t,\ldots,g_{\lambda_j}(x_n)^t\big)^t
\in \mathbb{R}^{n\times p},
\qquad 
\epsilon_{\lambda_j}
:= (\epsilon_{1,\lambda_j},\ldots,\epsilon_{n,\lambda_j})^t.
\]
We assume $g_{\lambda_j,0}(x_i)=0$. If $g_{\lambda_j,0}(x_i)$ is non-zero, this term can be subtracted from the left-hand side of Eq.~\eqref{eq15}, as illustrated in the numerical examples of Section~\ref{sec:num}.
Note that the same experimental vector $z$ appears in all $m$ equations, 
as it is compared to the model output evaluated at each $\lambda_j$. 
Accordingly, $z$ is replicated $m$ times in the matrix formulation below, which gathers these $m$ equations into a single matrix expression:
\begin{equation}
     (z,\ldots,z)
     =
     \big(g_{\lambda_1}(x)\theta(\lambda_1),\ldots,g_{\lambda_m}(x)\theta(\lambda_m)\big)
     + (\epsilon_{\lambda_1},\ldots,\epsilon_{\lambda_m}).
     \label{eq_mat}
\end{equation}

Let $\mathbf{z} := (z,\ldots,z) \in \mathbb{R}^{n\times m}$ be the matrix of the $m$ copies of $z$, 
and let the associated macro-parameter matrix be 
$\Theta_m := (\theta(\lambda_1),\ldots,\theta(\lambda_m))^t \in \mathbb{R}^{m\times p}$.  
Each $\theta(\lambda_j)$ follows a multivariate normal distribution arising from Eq.~\eqref{eq12}.  
One way to infer $\Theta_m$ is to work with its vectorized form, denoted 
$\vec{\Theta}_m := \mathrm{vec}(\Theta_m)$ \citep{barratt2018matrix} 
(see Appendix~\ref{vec}), whose prior density is written as:
\begin{equation}
    \pi(\vec{\Theta}_m | \phi)
    \propto |\C_{\phi}|^{-1/2}
    \exp\!\left\{
    -\frac12 \big(\vec{\Theta}_m-\vec{M}_\beta \big)^t 
    \C_{\phi}^{-1}\big(\vec{\Theta}_m-\vec{M}_\beta \big)
    \right\},
    \label{eq17}
\end{equation}
where
\begin{equation}
    \vec{M}_\beta 
    = \big(
        m_{\beta_1}(\lambda_1),\ldots,m_{\beta_p}(\lambda_1),\,
        \ldots,\,
        m_{\beta_1}(\lambda_m),\ldots,m_{\beta_p}(\lambda_m)
      \big)^t
    \in \mathbb{R}^{pm},
\end{equation}
and
\begin{equation}
    \C_{\phi} = \{\cov(\theta(\lambda_j),\theta(\lambda_{j'}))\}_{j,j'=1}^m,
    \qquad 
    \cov(\theta(\lambda_j),\theta(\lambda_{j'}))
    = \mathrm{diag}\big\{ \sigma_l^2 K_{\psi_l}(\lambda_j,\lambda_{j'}) \big\}_{l=1}^p,
\end{equation}
and
\begin{equation}
\phi := \{ (\beta_l, \sigma_l^2, \psi_l) \}_{l=1}^p.
\end{equation}
Although the parameters $\beta_l$ enter only through the prior mean $\vec{M}_\beta$, we collect them together with the covariance hyperparameters $(\sigma_l^2,\psi_l)$ into the vector $\phi$ for notational convenience.

\medskip
Let $\vec{\mathbf{z}} := \mathrm{vec}(\mathbf{z})$ denote the vectorized form of the matrix 
$\mathbf{z} = (z,\ldots,z)$, and let $\vec{\epsilon}$ denote the vectorized form of 
$(\epsilon_{\lambda_1},\ldots,\epsilon_{\lambda_m})$. One can then rewrite 
Eq.~\eqref{eq_mat} in its vectorized form:
\begin{equation}
\begin{aligned}
    \vec{\mathbf{z}}
    &= 
    \begin{pmatrix}
        g_{\lambda_1}(x)\,\theta(\lambda_1) \\[1mm]
        \vdots \\[1mm]
        g_{\lambda_m}(x)\,\theta(\lambda_m)
    \end{pmatrix}
    + \vec{\epsilon} \\[4mm]
    &=\; G\,\vec{\Theta}_m + \vec{\epsilon},
    \qquad 
    \vec{\epsilon} \sim \mathcal{N}\!\left(0,\,\Sigma_{\vec{\epsilon}} \right),
\end{aligned}
\end{equation}
where
\begin{equation}
    G = \mathrm{diag}\big(g_{\lambda_1}(x),\ldots,g_{\lambda_m}(x)\big) 
    \in \mathbb{R}^{nm\times pm},
\end{equation}
\begin{equation}
    \Sigma_{\vec{\epsilon}} 
    = \mathrm{diag}\big(\Sigma_{\epsilon_{\lambda_1}},\ldots,\Sigma_{\epsilon_{\lambda_m}}\big)
    \in \mathbb{R}^{nm\times nm},
\end{equation}
\begin{equation}
\Sigma_{\epsilon_{\lambda_j}}
= \mathrm{diag}\!\left(
\sigma_{\epsilon_1}^2 + \delta_{\lambda_j,1}^2,\;
\ldots,\;
\sigma_{\epsilon_n}^2 + \delta_{\lambda_j,n}^2
\right),
\qquad 1\le j\le m.
\end{equation}
Assume that the covariance matrix $\Sigma_{\vec{\epsilon}}$ is symmetric
positive definite. The likelihood of $\vec{\mathbf{z}}$ conditionally on $\vec{\Theta}_m$ is given by:
\begin{equation}
    \lik(\vec{\mathbf{z}}|\vec{\Theta}_m)
    \propto 
    \exp\!\left\{
    -\frac{1}{2} 
    \left(\vec{\mathbf{z}} - G\vec{\Theta}_m \right)^t 
    \Sigma^{-1}_{\vec{\epsilon}}
    \left(\vec{\mathbf{z}} - G\vec{\Theta}_m \right)
    \right\}.
    \label{eq18}
\end{equation}
Finally, by Bayes' formula:
\begin{equation}
\pi(\vec{\Theta}_m|\vec{\mathbf{z}},\phi)
\propto 
\lik(\vec{\mathbf{z}}| \vec{\Theta}_m)\,
\pi(\vec{\Theta}_m|\phi).
\end{equation}

\begin{theorem}\label{theo1}
Assume that the prior covariance matrix $\C_{\phi}$ is symmetric
positive definite. We can define
\begin{equation}
\Delta^{-1}:=G^t\Sigma_{\vec{\epsilon}}^{-1}G
=
\mathrm{diag}\!\Big(
g_{\lambda_1}(x)^t\Sigma_{\epsilon_{\lambda_1}}^{-1}g_{\lambda_1}(x),
\ldots,
g_{\lambda_m}(x)^t\Sigma_{\epsilon_{\lambda_m}}^{-1}g_{\lambda_m}(x)
\Big).
\end{equation}
The posterior distribution \(\pi(\vec{\Theta}_m|\vec{\mathbf{z}},\phi)\) 
is multivariate normal with mean \(\mu_\phi\) and covariance matrix \(\Sigma_\phi\) given by:
\begin{equation}
    \mu_\phi 
    =
    \Sigma_\phi
    \big(\C_{\phi}^{-1}\vec{M}_\beta + G^{t}\Sigma_{\vec{\epsilon}}^{-1}\vec{\mathbf{z}}\big)
    \in \mathbb{R}^{pm},
\end{equation}
\begin{equation}
    \Sigma_\phi 
    =
    \big(\Delta^{-1} + \C_{\phi}^{-1}\big)^{-1}
    \in \mathbb{R}^{pm \times pm}.
\end{equation}
See Appendix~\ref{B.1} for the proof.
\end{theorem}

\subsection{Predictive distribution of $\theta(\lambda)$}
\label{sec:pred}

For any new set of realizations $\lambda^\star=(\lambda_1^\star,\ldots,\lambda_k^\star)^t$ drawn from $\pi(\lambda|w)$, the predictive distribution of
\begin{equation}
\vec{\theta}(\lambda^\star)
:=\text{vec}\Big(
\big(\theta(\lambda_1^\star)^t,\ldots,\theta(\lambda_k^\star)^t\big)^t\Big)
\in\mathbb{R}^{pk}
\end{equation}
is obtained by integrating the conditional Gaussian distribution
$\pi(\vec{\theta}(\lambda^\star)|\vec{\Theta}_m,\phi)$

\begin{equation}
\pi_{\mathrm{pred}}\big(\vec{\theta}(\lambda^\star)| \vec{\mathbf{z}}, \phi\big)
= \int_{\mathcal{T}^m} 
\pi\big(\vec{\theta}(\lambda^\star)|\vec{\Theta}_m,\phi\big)\,
\pi\big(\vec{\Theta}_m|\vec{\mathbf{z}}, \phi\big)\, d\vec{\Theta}_m.
\end{equation}

\begin{theorem}\label{theo2}
Let $\lambda^{\star\prime}=(\lambda_1^{\star\prime},\ldots,\lambda_{k'}^{\star\prime})^t$. The predictive distribution
$\pi_{\mathrm{pred}}(\vec{\theta}(\lambda^\star)|\vec{\mathbf{z}},\phi)$
is a multivariate normal distribution with mean
$\bar{\theta}_{\mathrm{pred}}(\lambda^\star)$ and cross-covariance matrix
$\mathbf{\Sigma}_{\mathrm{pred}}(\lambda^\star,\lambda^{\star\prime})$ given by
\begin{equation}
\label{eq:pred_mean}
\bar{\theta}_{\mathrm{pred}}(\lambda^\star)
=
\vec{m}_\beta(\lambda^\star)
+
\C(\lambda^\star,D_m)\,
\C_{\phi}^{-1}
\big(\mu_\phi-\vec{M}_\beta\big)
\in\mathbb{R}^{pk},
\end{equation}
and
\begin{equation}
\label{eq:pred_cov}
\mathbf{\Sigma}_{\mathrm{pred}}(\lambda^\star,\lambda^{\star\prime})
=
\mathbf{\Sigma}_{\mathrm{cond}}(\lambda^\star,\lambda^{\star\prime})
+
\C(\lambda^\star,D_m)\,
\C_{\phi}^{-1}
\mathbf{\Sigma}_\phi
\C_{\phi}^{-1}\,
\C(D_m,\lambda^{\star\prime})
\in\mathbb{R}^{pk\times pk^{\prime}},
\end{equation}
where $(\mu_\phi,\mathbf{\Sigma}_\phi)$ are the posterior mean and covariance
of $\vec{\Theta}_m$ given in Theorem~\ref{theo1}, and
\begin{align}
    &\vec{m}_\beta(\lambda^\star)
    =
    \begin{pmatrix}
    m_{\beta_1}(\lambda_1^\star), 
    \ldots,
    m_{\beta_p}(\lambda_1^\star),
    \ldots,
    m_{\beta_1}(\lambda_k^\star),
    \ldots,
    m_{\beta_p}(\lambda_k^\star)
    \end{pmatrix}^t
    \in \mathbb{R}^{pk},
    \\[0.5em]
   &\C(\lambda^\star,D_m)
   =
   \Big[
   \operatorname{Cov}\big(\theta(\lambda_i^\star), \theta(\lambda_j)\big)
   \Big]_{1\le i\le k,\;1\le j\le m}
   \in \mathbb{R}^{pk \times pm},
   \\[0.5em]
   &\mathbf{\Sigma}_{\mathrm{cond}}(\lambda^\star,\lambda^{\star\prime})
   =
   \C(\lambda^\star,\lambda^{\star\prime})
   -
   \C(\lambda^\star,D_m)\,
   \C_{\phi}^{-1}\,
   \C(D_m,\lambda^{\star\prime})
   \in \mathbb{R}^{pk \times pk^{\prime}}.
\end{align}
Each entry of $\C(\lambda^\star,D_m)$ and
$\C(\lambda^\star,\lambda^{\star\prime})$ is a $p\times p$ covariance
block.
See Appendix~\ref{B.2} for the proof.
\end{theorem}

The expression of $\bar{\theta}_{\mathrm{pred}}(\lambda^\star)$ follows the classical form of the conditional Gaussian process mean, except that the latent vector $\vec{\Theta}_m$ is replaced by its posterior expectation $\mu_\phi$. An important advantage of GP--LinCC is that $\bar{\theta}_{\mathrm{pred}}(\lambda^\star)$ provides a predictor of $\theta(\lambda^\star)$ without requiring the regression vectors $g_{\lambda^\star}(x_i)$ for $1\leq i\leq n$. The predictive covariance $\Sigma_{\mathrm{pred}}(\lambda^\star,\lambda^{\star\prime})$ consists of two parts.
The first component, $\Sigma_{\mathrm{cond}}(\lambda^\star,\lambda^{\star\prime})$, is the usual GP interpolation variance, that is, the conditional covariance one would obtain as if the values of $\vec{\Theta}_m$ were exactly known.
It governs the transfer of uncertainty induced by Gaussian process interpolation across $\lambda$.
The second term captures the posterior uncertainty on $\vec{\Theta}_m$ and is governed by $\Sigma_\phi$. This term propagates the posterior uncertainty on the unknown parameters $\vec{\Theta}_m$ to unseen values of $\lambda$ through the covariance function.
As $m$ increases, only $\Sigma_{\mathrm{cond}}(\lambda^\star,\lambda^\star)$ is structurally reduced, because a richer design $D_m$ limits the possible excursions of the GP between design points.
By contrast, the second component of the predictive variance, driven by the uncertainty on $\vec{\Theta}_m$ through $\Sigma_\phi$, is primarily decreased by enlarging the experimental sample size $n$.

For notational simplicity, we omit the superscript “$\star$” in the predictive formulas and write $\theta(\lambda)$ instead of $\theta(\lambda^\star)$ in the remainder of the paper.


\medskip
All the previous formulas depend on the hyperparameters $\phi$, which are not known a priori. We estimate them using an empirical Bayes procedure based on marginal likelihood maximization \citep{rasmussen2006gaussian}.
This approach maximizes the marginal likelihood with respect to $\phi$, obtained by integrating out $\vec{\Theta}_m$ from $\lik(\vec{\mathbf{z}}|\vec{\Theta}_m)$:
\begin{equation}
\widehat{\phi}
=\argmax_{\phi}\;
\int_{\mathcal{T}^m}
\lik(\vec{\mathbf{z}}|\vec{\Theta}_m)\,
\pi(\vec{\Theta}_m|\phi)\,d\vec{\Theta}_m.
\label{eq25}
\end{equation}
Appendix \ref{B.4} demonstrates that this integral has a closed form. 

\medskip
When the true functional parameter $\theta_{\text{true}}(\lambda)$ is known, as in numerical experiments, parameter recovery can be evaluated using the predictive Integrated Mean-Squared Error (IMSE). It is defined as the posterior predictive squared loss relative to $\theta_{\text{true}}(\lambda)$, averaged with respect to $\pi(\lambda|w)$:
\begin{equation}
\label{eq::IMSE}
\text{IMSE}
=
\int_{\Lambda}
\Exp\Big[
\left(\theta(\lambda)-\theta_{\text{true}}(\lambda)\right)^{t}
\left(\theta(\lambda)-\theta_{\text{true}}(\lambda)\right)
|
\vec{\mathbf{z}},\phi
\Big]\,
\pi(\lambda|w)\,d\lambda
\end{equation}
where the inner expectation expands as
\begin{equation}
\Exp\!\Big[
(\theta(\lambda)-\theta_{\text{true}}(\lambda))^{t}
(\theta(\lambda)-\theta_{\text{true}}(\lambda))
|
\vec{\mathbf{z}},\phi
\Big]
=
\sum_{u=1}^{p}
\Big(
\Sigma_{\mathrm{pred}}(\lambda,\lambda)_{u,u}
+
\left(\bar{\theta}_{\mathrm{pred}, u}(\lambda)-\theta_{\text{true},u}(\lambda)\right)^2
\Big).
\end{equation}
In practice, the IMSE is estimated by Monte Carlo from a sample of size \(N_\lambda\):
\begin{align}
\text{IMSE}
\approx 
\frac{1}{N_{\lambda}}
\sum_{j=1}^{N_{\lambda}}
\sum_{u=1}^{p}
\Big(
\Sigma_{\mathrm{pred}}(\lambda_j,\lambda_j)_{u,u}
+
(\bar{\theta}_{\mathrm{pred},u}(\lambda_j)
     -\theta_{\mathrm{true},u}(\lambda_j))^2
\Big).
\end{align}
Then, the predictive cut distribution is defined as
\begin{equation}
\label{eq:pred-cut}
\pi_{\mathrm{cut},\mathrm{pred}}(\lambda,\theta|w,\vec{\mathbf{z}},\phi)
:=\pi(\lambda|w)\,
\pi_{\mathrm{pred}}(\theta(\lambda)|\vec{\mathbf{z}},\phi).
\end{equation}
Eq.~\eqref{eq:pred-cut} mirrors the structure of the cut distribution in Eq.~\eqref{eq4}, with the density $\pi(\theta|\lambda,z)$ replaced by the predictive density $\pi_{\mathrm{pred}}(\theta(\lambda)|\vec{\mathbf{z}},\phi)$. In the synthetic examples presented in Section \ref{sec:num}, we will examine the adequacy of this predictive approximation with respect to the target density $\pi_{\mathrm{cut}}(\lambda,\theta|w,z)$. In these examples, the linear model $g_{\lambda}(x_i)$ in Eq.~\eqref{eq13} is known for all $\lambda$, which allows the explicit computation, for any $\lambda$, of the target conditional posterior distribution of $\theta$ given $\lambda$. Under a Gaussian prior for $\theta|\lambda$,
\begin{equation}
\theta(\lambda)
\sim
\mathcal{N}\big(m_0(\lambda),\, V_0(\lambda)\big),
\end{equation}
we have
\begin{equation}
\theta|\lambda,z
\sim
\mathcal{N}\big(m_{\mathrm{post}}(\lambda),\, V_{\mathrm{post}}(\lambda)\big),
\end{equation}
with
\begin{equation}
V_{\mathrm{post}}(\lambda)^{-1}
=
V_0(\lambda)^{-1}
+
g_{\lambda}(x)^t
\Sigma_{\epsilon_{\lambda}}^{-1}
g_{\lambda}(x),
\end{equation}
and
\begin{equation}
m_{\mathrm{post}}(\lambda)
=
V_{\mathrm{post}}(\lambda)
\Big(
V_0(\lambda)^{-1} m_0(\lambda)
+
g_{\lambda}(x)^t
\Sigma_{\epsilon_{\lambda}}^{-1}
z
\Big).
\end{equation}
Specifying instead the Jeffreys prior,
\begin{equation}
\pi(\theta|\lambda) \propto 1,
\end{equation}
leads to
\begin{equation}
\label{eq::target}
\theta|\lambda,z
\sim
\mathcal{N}_p\!\left(
\widehat{\theta}(\lambda),
\left(g_{\lambda}(x)^t \Sigma_{\epsilon_{\lambda}}^{-1}
      g_{\lambda}(x)\right)^{-1}
\right),
\end{equation}
where
\begin{equation}
\widehat{\theta}(\lambda)
=
\left(g_{\lambda}(x)^t\Sigma_{\epsilon_{\lambda}}^{-1}
      g_{\lambda}(x)\right)^{-1}
g_{\lambda}(x)^t\,\Sigma_{\epsilon_{\lambda}}^{-1} z.
\end{equation}
The discrepancy between the predictive and target conditional distributions may decrease once the design $D_m$
becomes sufficiently informative. This occurs when $m$ is large enough for the
conditional covariance term $\Sigma_{\mathrm{cond}}(\lambda,\lambda)$ to remain
uniformly small over the support of $\lambda$. In this regime, the interpolation
uncertainty is largely suppressed, and the predictive variance induced by
Eq.~\eqref{eq:pred_cov} is dominated by the propagation of posterior uncertainty
through the term
\begin{equation}
\C(\lambda,D_m)\C_\phi^{-1}\Sigma_\phi\C_\phi^{-1}\C(D_m,\lambda),
\end{equation}
leading to a closer agreement with the target posteriors. This second contribution does not structurally decrease with $m$
for fixed hyperparameters. However, when $\phi$ is estimated by empirical Bayes,
the increase of $m$ may induce an additional contraction of $\Sigma_\phi$. In that case, the predictive
variance may become substantially smaller than the target posterior variance
$V_{\mathrm{post}}(\lambda)$, resulting in an over-concentrated predictive conditional distribution.

\subsection{Predictive behavior of the chained model}

For a given input configuration $x_i$, we define
\begin{equation}
r_i(\lambda)
:= g_{\lambda}(x_i)^t \theta(\lambda).
\end{equation}
The predictive distribution of the output of the calibrated model is given by
\begin{equation}
\label{predictive}
  r_i(\lambda)
  \;\big|\; \vec{\mathbf{z}},\phi
  \;\sim\;
  \mathcal{N}\!\left(
  g_{\lambda}(x_i)^t\,\bar{\theta}_{\mathrm{pred}}(\lambda),\;
  g_{\lambda}(x_i)^t\,\Sigma_{\mathrm{pred}}(\lambda,\lambda)\,
  g_{\lambda}(x_i)
  \right).
\end{equation}
Under a cross-validation scheme, Eq.~\eqref{predictive} can be reformulated in a leave-one-out setting. For $1 \leq i \leq n$,
\begin{equation}
\label{eq33}
\begin{aligned}
  r_i(\lambda)
  \;\big|\; \vec{\mathbf{z}}_{-i},\phi
  &\;\sim\;
  \mathcal{N}\!\left(
      g_{\lambda}(x_i)^t \bar{\theta}_{\mathrm{pred},-i}(\lambda),\,
      \right.\\[1mm]
  &\qquad\qquad\left.
      g_{\lambda}(x_i)^t
      \Sigma_{\mathrm{pred},-i}(\lambda,\lambda)
      g_{\lambda}(x_i)
  \right).
\end{aligned}
\end{equation}
where $\bar{\theta}_{\mathrm{pred},-i}$ and $\Sigma_{\mathrm{pred},-i}$ are obtained from
Theorem~\ref{theo2} by replacing $\vec{\mathbf{z}}$ with $\vec{\mathbf{z}}_{-i}$
(i.e.\ $\vec{\mathbf{z}}$ with the $i$th observation removed). In the following, we consider the predictive random variable
$r_i$ obtained after marginalizing over $\lambda\sim\pi(\lambda|w)$. The predictive mean of $r_i$, obtained after marginalizing over 
$\lambda\sim\pi(\lambda|w)$, follows from the law of total expectation and is given by
\begin{equation}
\label{eq:mean-Ri}
\mathbb{E}\big[r_i \vert \vec{\mathbf{z}}_{-i},\phi\big]
=
\mathbb{E}_{\lambda\vert w}\!\Big[
g_{\lambda}(x_i)^t\,\bar{\theta}_{\mathrm{pred},-i}(\lambda)
\Big].
\end{equation}
In order to quantify the total predictive uncertainty at $x_i$,
the variance of $r_i$ can then be decomposed with respect to
$\lambda$ using the law of total variance as
\begin{align}
\label{eq:var-total-Ri}
\Var\big(r_i \vert \vec{\mathbf{z}}_{-i},\phi\big)
&=
\mathbb{E}_{\lambda\vert w}\!\Big[
    \Var\big(r_i(\lambda)\vert \vec{\mathbf{z}}_{-i},\phi\big)
\Big]
+
\Var_{\lambda\vert w}\!\Big(
    \mathbb{E}\big[r_i(\lambda)\vert \vec{\mathbf{z}}_{-i},\phi\big]
\Big)
\\[2mm]
\label{eq:var-decomp}
&=
\mathbb{E}_{\lambda\vert w}\!\Big[
    g_{\lambda}(x_i)^t\,
    \Sigma_{\mathrm{pred},-i}(\lambda,\lambda)\,
    g_{\lambda}(x_i)
\Big]
+
\Var_{\lambda\vert w}\!\Big(
    g_{\lambda}(x_i)^t\,\bar{\theta}_{\mathrm{pred},-i}(\lambda)
\Big).
\end{align}
The first term in Eq.~\eqref{eq:var-decomp} corresponds to the predictive variance
of $r_i(\lambda)$ conditionally on $\lambda$, averaged with respect to
$\pi(\lambda|w)$. It therefore quantifies the residual posterior uncertainty on
$\theta(\lambda)$ after calibration, projected onto the scalar output $r_i$,
for each fixed value of $\lambda$.
The second term represents the variance, under $\pi(\lambda|w)$, of the
conditional predictive mean
$g_{\lambda}(x_i)^t\,\bar{\theta}_{\mathrm{pred},-i}(\lambda)$, and thus measures
the sensitivity of the calibrated prediction at $x_i$ to the uncertainty on
$\lambda$. 

When the mean and variance of the conditional predictive distributions in
Eq.~\eqref{eq33} remain similar for different values of $\lambda$, and provided
that the full posterior $\pi(\lambda|\vec{\mathbf{z}}_{-i},w)$ does not place
substantial mass outside the support of $\pi(\lambda|w)$, the predictive
distribution of $r_i$ becomes weakly sensitive to the choice of the marginal
distribution of $\lambda$ and is therefore expected to be close to its full
counterpart, in which the mean and variance in
Eqs.~\eqref{eq:mean-Ri}–\eqref{eq:var-decomp} are taken with respect to
$\pi(\lambda|\vec{\mathbf{z}}_{-i},w)$.
This situation is referred to as a \emph{compensation effect}, and typically
occurs when distinct pairs $(\lambda_1,\theta(\lambda_1))$ and
$(\lambda_2,\theta(\lambda_2))$ lead to similar likelihood values, i.e.\ when the
downstream model is non-identifiable.

To provide a more detailed characterization of such a potential compensation
effect, it is therefore useful to examine how the predictive distribution varies
with $\lambda$. If a compensation effect occurs to some extent, then for typical values
of $\lambda_1 \neq \lambda_2$ drawn from $\pi(\lambda|w)$, the corresponding
predictive densities
\begin{equation}
\label{eq:pred-two}
\pi\!\left(
    r_i(\lambda_1)
    \,\big|\,
    \vec{\mathbf{z}}_{-i},\phi
\right)
\quad\text{and}\quad
\pi\!\left(
   r_i(\lambda_2)
    \,\big|\,
    \vec{\mathbf{z}}_{-i},\phi
\right)
\end{equation}
are expected to be similar. Such a compensation effect arises when the downstream model is non-identifiable to some extent, 
meaning that distinct pairs $(\lambda_1,\theta(\lambda_1))$ and 
$(\lambda_2,\theta(\lambda_2))$ lead to the same likelihood 
(see \citep{cole2020parameter} for an in-depth discussion of nonidentifiability). Therefore, the predictive credibility interval of the 
random variable
\begin{equation}
\label{eq:diff-predictive}
r_i(\lambda_1)
-
r_i(\lambda_2)
\;\big|\;
\vec{\mathbf{z}}_{-i},\phi
\end{equation}
is likely to contain\, $0$.
This random variable is Gaussian with mean
\begin{equation}
\mu_i(\lambda_1,\lambda_2):=
\begin{pmatrix}
g_{\lambda_1}(x_i) \\
-\,g_{\lambda_2}(x_i)
\end{pmatrix}^{t}
\begin{pmatrix}
\bar{\theta}_{\mathrm{pred},-i}(\lambda_1) \\
\bar{\theta}_{\mathrm{pred},-i}(\lambda_2)
\end{pmatrix},
\end{equation}
and variance
\begin{equation}
\sigma_i^2(\lambda_1,\lambda_2)
:=
\begin{pmatrix}
g_{\lambda_1}(x_i) \\
-\,g_{\lambda_2}(x_i)
\end{pmatrix}^{t}
\begin{pmatrix}
\Sigma_{\mathrm{pred},-i}(\lambda_1,\lambda_1)
&
\Sigma_{\mathrm{pred},-i}(\lambda_1,\lambda_2)
\\[1mm]
\Sigma_{\mathrm{pred},-i}(\lambda_2,\lambda_1)
&
\Sigma_{\mathrm{pred},-i}(\lambda_2,\lambda_2)
\end{pmatrix}
\begin{pmatrix}
g_{\lambda_1}(x_i) \\
-\,g_{\lambda_2}(x_i)
\end{pmatrix}.
\end{equation}
We can compute an empirical coverage probability for $0$ at level $1-\alpha$
using $N$ i.i.d.\ sample pairs $(\lambda_1,\lambda_2)\sim \pi(\lambda|w)\times \pi(\lambda|w)$:
\begin{equation}
\hat{\Delta}(\alpha,x_i)
=
\frac{1}{N}
\sum_{j=1}^N
\mathbf{1}_{\left\{
0 \in 
\left[
\mu_i(\lambda_{1,j},\lambda_{2,j})
\;\pm\;
q_{1-\alpha/2}\,
\sqrt{\sigma^2_i(\lambda_{1,j},\lambda_{2,j})}
\right]
\right\}},
\label{eq34}
\end{equation}
where $q_{1-\alpha/2}$ denotes the $(1-\alpha/2)$ quantile of the standard Gaussian distribution. One expects $\hat{\Delta}(\alpha,x_i)$ to be close to $1-\alpha$. When $\alpha=5\%$, the presence of a
compensation effect should be questioned whenever $\hat{\Delta}(5\%,x_i)$ is significantly below $95\%$.


Investigating the presence of a compensation effect may justify the use of the cut approach rather than the full approach, even in situations where the latter is theoretically optimal, i.e., when the modeling assumptions of Eq.~\eqref{eq2} are fully satisfied (no model discrepancy and Gaussian experimental uncertainty). In this well-specified framework, although the cut posterior differs from the full posterior, this discrepancy is expected to have only a limited impact on the predictive distribution of the chained model when a compensation effect is present.


\section{Numerical examples}
\label{sec:num}
\subsection{One-dimensional examples}
To illustrate the performance of the GP--LinCC method, we start with the example in Eq.~(\ref{toy_example}) introduced in Section~\ref{cut_vs_full_simple_example}. Recall that the posterior distribution $\pi(\lambda|w)$ is Gaussian, $\mathcal{N}(\mu_{\lambda,\mathrm{cut}}, \sigma^2_{\lambda,\mathrm{cut}})$, with mean and variance given by
\begin{equation}
\mu_{\lambda,\mathrm{cut}} = \bar{w}, \qquad 
\sigma^2_{\lambda,\mathrm{cut}} = \frac{\sigma^2_w}{n_1}.
\end{equation}
where $n_1 = 15$ denotes the size of the vector $w$.

\medskip
The functional parameter $\theta(\lambda)\in\mathbb{R}$ is assumed to follow a GP with a constant mean function $m_{\beta}(\lambda)=\beta$ and a Matérn $5/2$ covariance function given by Eq.~\eqref{matern}. The quantities $g_{\lambda,0}(x_i)$ and $g_{\lambda,1}(x_i)$ introduced in Section~\ref{sec:gplincc} are therefore $x_i\lambda$ and $1$, respectively. Note that, by construction of GP--LinCC, the predictive distribution $\pi_{\mathrm{pred}}(\theta(\lambda)|\vec{\zgris},\phi)$ interpolates the posterior mean of $\theta|\lambda$, which is
\begin{equation}
   \mu_{\theta|\lambda} := \bar{z} - \bar{x}\,\lambda. 
\label{eq53S5}
\end{equation}
Since Eq.~\eqref{eq53S5} forms a perfectly linear function of 
$\lambda$, interpolating it with a smooth Matérn~5/2 GP leads to an almost 
collinear covariance matrix, causing numerical conditioning issues \citep{PengWu2014} 
and making the resulting GP--LinCC predictive distribution unreliable. To address 
this, we apply the bounded, smooth reparameterization $T:\mathbb{R}\to]-1,1[$, which 
moves the linear posterior means onto a non-linear scale on which the covariance 
matrix is well conditioned:
\begin{equation}
\label{trans_T}
\begin{aligned}
T(a) 
&= \frac{2}{1+\mathrm{e}^{-a}} - 1
= \frac{\mathrm{e}^{a}-1}{\mathrm{e}^{a}+1}
= \tanh\!\left(\frac{a}{2}\right),
&\qquad & a \in \mathbb{R},\\[2mm]
T^{-1}(b) 
&= \log\!\left(\frac{1+b}{\,1-b\,}\right)
= 2\,\mathrm{arctanh}(b),
&\qquad & b \in ]-1,1[.
\end{aligned}
\end{equation}
We can apply the GP--LinCC approach to the transformed paramter $\eta(\lambda):= T(\theta(\lambda))$. The design $D_m$ is constructed by mapping an LHS $m$-sample on $[0,1]$ onto $\pi(\lambda|w)$ via the inverse standard normal CDF.
Let $\widetilde{\pi}_{\mathrm{pred}}(\eta(\lambda)|\vec{\mathbf{z}},\phi)$
denote the resulting predictive distribution of $\eta(\lambda)$.
By a change of variables through $T^{-1}$, the induced predictive
distribution of $\theta(\lambda)$ is obtained by applying an $\mathrm{arctanh}$ transformation to a Gaussian predictive law, which yields a
non-standard distribution that can be referred to as \emph{tanh-normal}:
\begin{equation}
\label{eq54s5}
\pi_{\mathrm{pred}}\!\left(\theta(\lambda)|\vec{\mathbf{z}}, \phi\right)
= 
\widetilde{\pi}_{\mathrm{pred}}\!\left(
\frac{\mathrm{e}^{\theta(\lambda)} - 1}{\mathrm{e}^{\theta(\lambda)} + 1}\big|\vec{\mathbf{z}}, \phi
\right)
\left(
\frac{2\,\mathrm{e}^{\theta(\lambda)}}{\left(\mathrm{e}^{\theta(\lambda)}+1\right)^2}
\right).
\end{equation}
The GP hyperparameters $\phi=(\beta,\psi, \sigma^2)$ are estimated by marginal likelihood maximization (see Eq.~\eqref{eq25}). 


\medskip

For each sampled pair $(\lambda_{1,j},\lambda_{2,j})$ in Eq.~\eqref{eq34}, the empirical coverage probability is computed as follows:
\begin{enumerate}
\item Draw $N_\theta$ samples from the GP--LinCC predictive distribution associated with
$\big(\eta(\lambda_{1,j}),\,\eta(\lambda_{2,j})\big).$

\item Apply $T^{-1}$ to these $N_\theta$ samples to obtain draws of 
$\big(\theta(\lambda_{1,j}),\,\theta(\lambda_{2,j})\big)$.

\item For each draw, compute 
\begin{equation}
\label{eq:draw_coverage}
Z_{ij}
:=
x_i(\lambda_{1,j}-\lambda_{2,j})
+
\theta(\lambda_{1,j})
-
\theta(\lambda_{2,j}).
\end{equation}

\item Compute the empirical quantiles $q^{ij}_{\alpha/2}$ and 
$q^{ij}_{1-\alpha/2}$ from the $N_\theta$ values of $Z_{ij}$.
\end{enumerate}
Finally,
\begin{equation}
\hat{\Delta}(\alpha,x_i)
=
\frac{1}{N}
\sum_{j=1}^N
\mathbf{1}_{\{\,0 \in [\,q_{\alpha/2}^{ij},\,q_{1-\alpha/2}^{ij}\,]\,\}}.
\label{eq53}
\end{equation}
For each fixed \(\lambda\), the model output is equal to
\begin{equation}
r_i(\lambda) := x_i\lambda + \theta(\lambda).
\end{equation}
Since the
predictive distribution of \(\theta(\lambda)\) is \emph{tanh-normal}, the
distribution of \(r_i(\lambda)=x_i\lambda+\theta(\lambda)\) is the corresponding
translated \emph{tanh-normal} distribution, with density
\begin{equation}
\label{eq54-def}
\pi(r_i(\lambda)\,|\,\vec{\zgris}_{-i},\phi)
=
\widetilde{\pi}_{\mathrm{pred}}\!\left(
\frac{e^{\,r_i(\lambda) - x_i\lambda}-1}{
      e^{\,r_i(\lambda) - x_i\lambda}+1}
\,\Big|\,
\vec{\zgris}_{-i},\phi
\right)
\left(
\frac{2\,e^{\,r_i(\lambda) - x_i\lambda}}{
      \big(e^{\,r_i(\lambda) - x_i\lambda}+1\big)^2}
\right).
\end{equation}
We consider the two settings introduced in Section~\ref{cut_vs_full_simple_example}:
\begin{itemize}
    \item Non-identifiable setting: $x_i = 5$ for $1 \leq i \leq n = 15$.
    \item Identifiable setting: $x_i \sim \mathcal{U}(-6,6)$ for $1 \leq i \leq n = 15$.
\end{itemize}
In each setting, GP--LinCC is applied as previously described, and the resulting
predictive cut distribution is compared with the target cut distribution obtained
under the Jeffreys prior $\pi(\theta|\lambda)\propto 1$.


\subsubsection{Non-identifiable setting}
\begin{figure}
    \centering
    \includegraphics[width=0.6\textwidth]{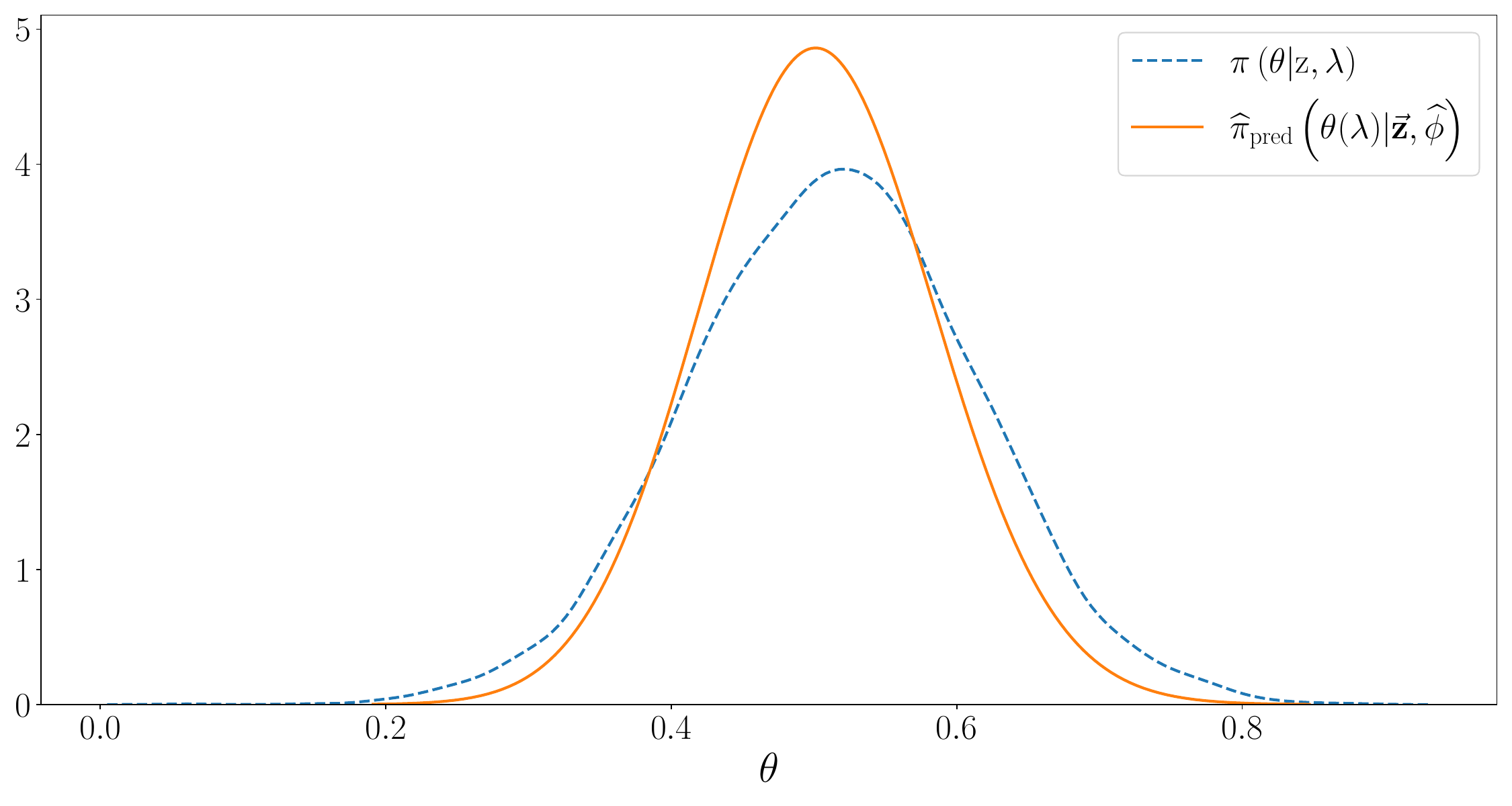}
\caption{Case $x_i = 5$ ($1 \leq i \leq n = 15$). Comparison between the target conditional density 
$\pi(\theta|z,\lambda)$ under the Jeffreys prior on $\theta$ and the tanh-normal density $\pi_{\mathrm{pred}}(\theta(\lambda)|\vec{\mathbf{z}},\hat{\phi})$ obtained using a design $D_m$ with 
$m=10$, for $\lambda = 1.238$. $\widehat{\pi}_{\mathrm{pred}}(\cdot|\cdot)$ denotes the KDE of $\pi_{\mathrm{pred}}(\cdot|\cdot)$.}

    \label{fig1results}
\end{figure}

\begin{figure}
    \centering
    \includegraphics[width=0.55\textwidth]{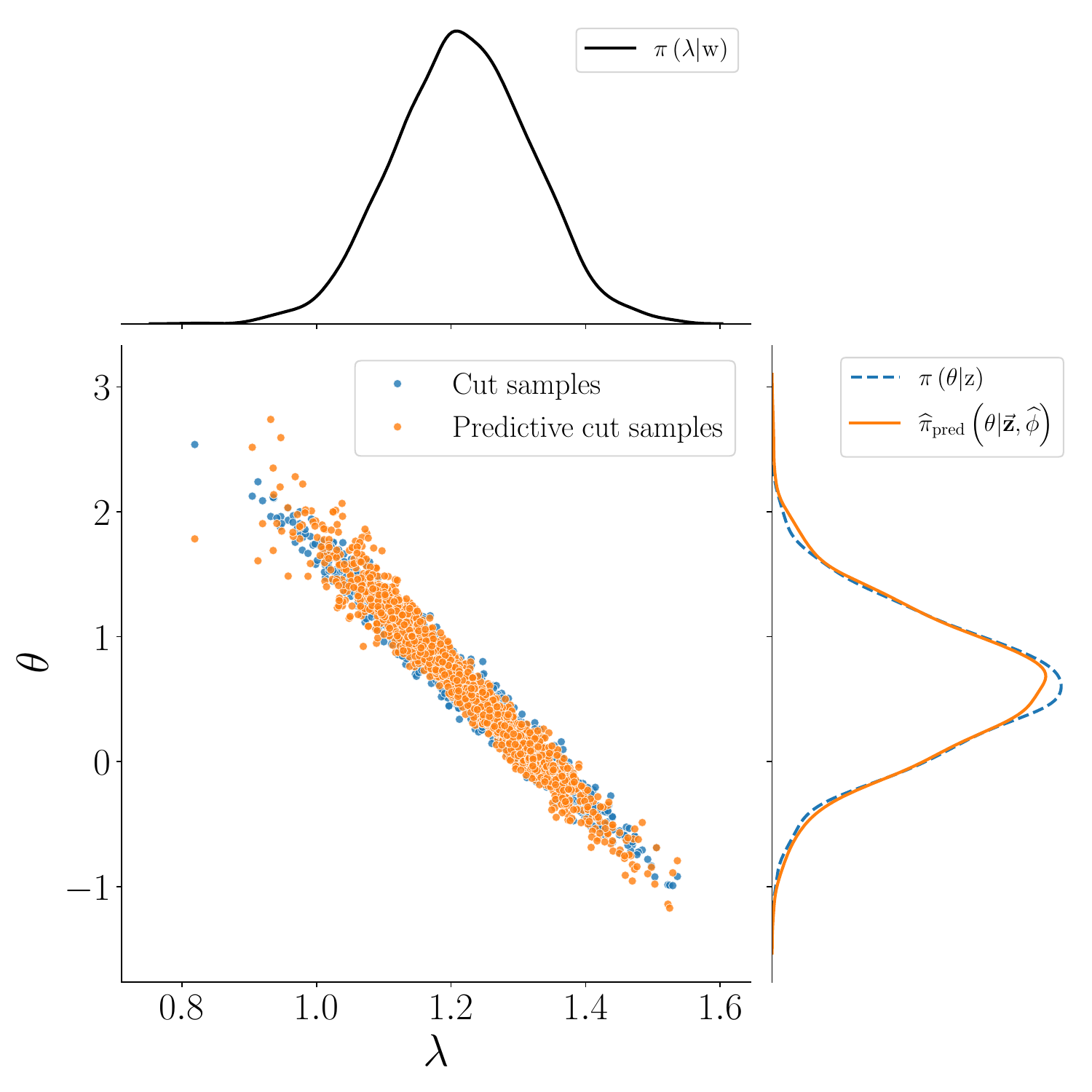}
\caption{Case $x_i = 5$ ($1 \leq i \leq n = 15$). Comparison between the target cut distribution under the 
Jeffreys prior on $(\lambda,\theta)$ and the GP--LinCC predictive cut distribution combined with $T^{-1}$, using a design $D_m$ of size $m=10$. For each distribution, $2000$ samples are displayed.}
    \label{fig2results}
\end{figure}

We constructed a design $D_m$ of size $m=10$. Figure~\ref{fig1results} displays a comparison between the KDE of the target conditional density \(\pi(\theta|z,\lambda)\) and the predictive tanh-normal density \(\pi_{\mathrm{pred}}(\theta(\lambda)|\vec{\zgris},\widehat{\phi})\) for the value \(\lambda=1.238\). The predictive density is close to the target conditional density, especially in terms of location, although it underestimates its dispersion. Similar behavior is observed for other draws of \(\lambda\sim \pi(\lambda|w)\).
Figure~\ref{fig2results} presents a comparison between the target cut distribution
given in Eq.~\eqref{eq4} and the predictive cut distribution obtained by replacing
\(\pi(\theta|z,\lambda)\) with \(\pi_{\mathrm{pred}}(\theta(\lambda)|\vec{\zgris},\widehat{\phi})\).
The two distributions show good agreement. However, the predictive cut distribution
is less accurate in the tails of the support of \(\lambda\). This occurs because \(\pi(\lambda|w)\) concentrates most of its mass around its mean, so the LHS-based design \(D_m\) contains fewer points in the tails.

\begin{figure}
    \centering
    \includegraphics[width=0.6\textwidth]{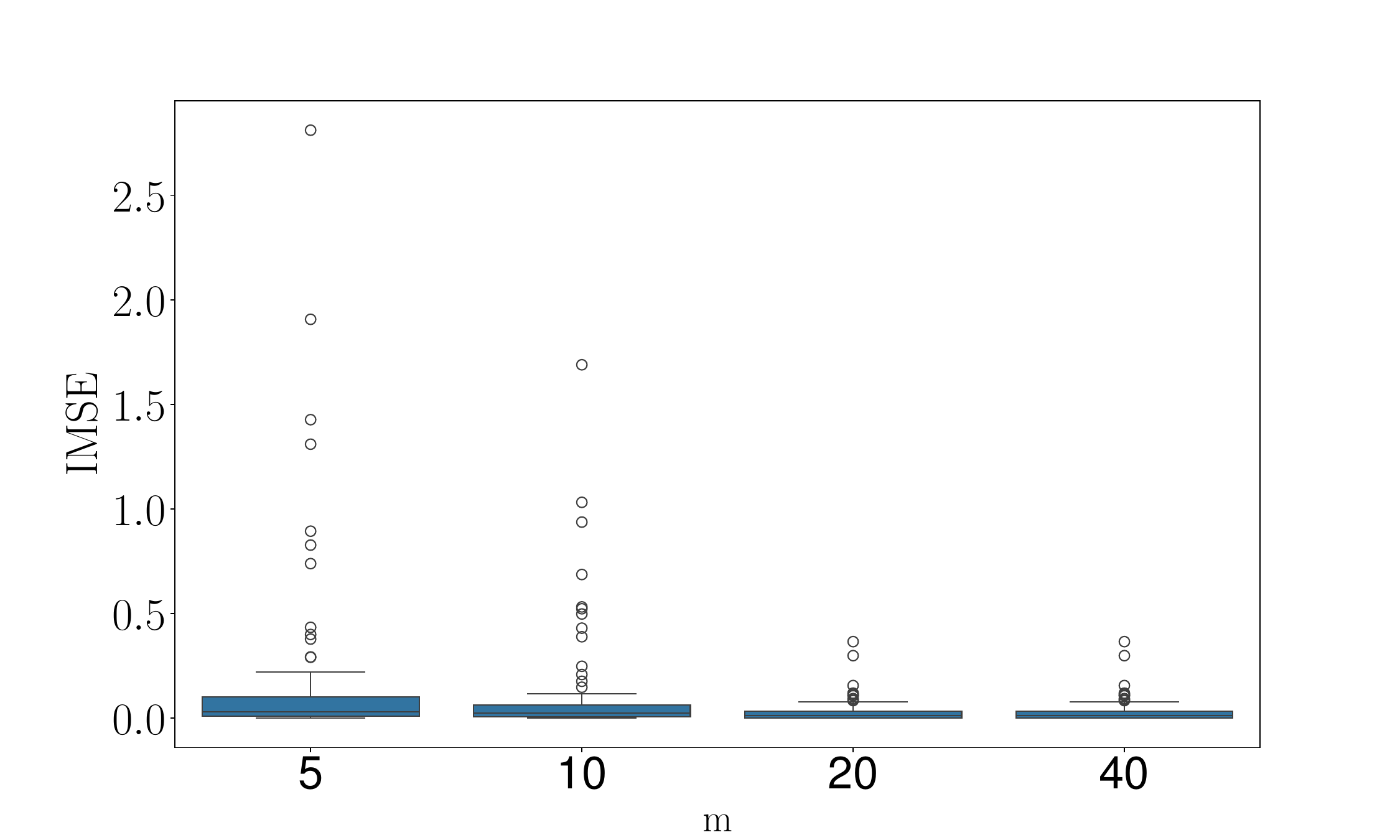}
\caption{Case $x_i = 5$ ($1 \leq i \leq n = 15$). Boxplot of the IMSE criterion for different sizes $m\in \{5, 10,20, 40 \}$ of $D_m$ and for a fixed sample $z$ of size $n=15$. Each boxplot is obtained from $1000$ samples.
}

    \label{figimse}
\end{figure}

In addition,  Figure \ref{figimse} shows the boxplots of the IMSE criterion for different sizes of $D_m$ and for a fixed sample $z$ of size $n = 15$. These boxplots, associated with each design, are obtained from $1000$ samples of the conditional distribution $\pi(\theta|z, \lambda)$ and the conditional distribution $\pi_{\mathrm{pred}}(\theta(\lambda)|\vec{\zgris},\widehat{\phi})$ provided  by GP--LinCC approach. As the size  $m$ of $D_m$ increases, the IMSE decreases, highlighting the good predictive ability of the GP--LinCC approach.

A compensation effect was expected due to the non-identifiability of the model. To assess this, the empirical coverage probabilities in Eq.~\eqref{eq53} were computed with \(\alpha = 5\%\), using \(N = 5000\) sample pairs \((\lambda_1,\lambda_2)\) generated from \(\pi(\lambda|w)\) and \(N_\theta = 5000\) samples of \(\theta\) drawn from the GP--LinCC predictive distribution at \((\lambda_1,\lambda_2)\). The results exceeded \(95\%\), confirming the presence of a compensation effect. 


\subsubsection{Identifiable setting}
\begin{figure}[ht!]
    \centering
    \includegraphics[width=0.6\textwidth]{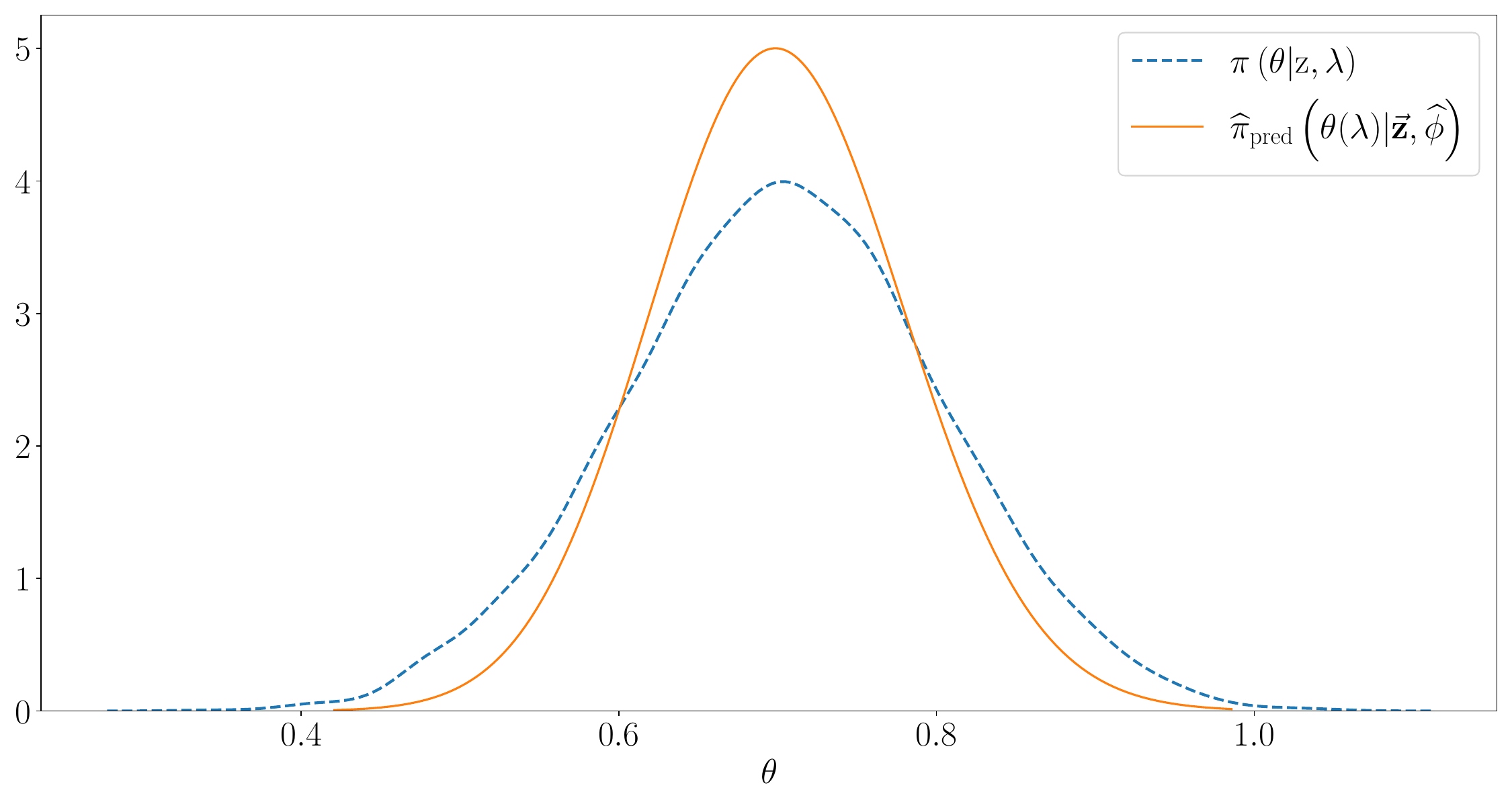}
\caption{Case $x_i \sim \mathcal{U}(-6,6)$ ($1 \leq i \leq n = 15$). Comparison between the target conditional density 
$\pi(\theta|z,\lambda)$ under the Jeffreys prior on $\theta$ and the tanh-normal density 
$\pi_{\mathrm{pred}}(\theta(\lambda)|\vec{\mathbf{z}},\hat{\phi})$ obtained using a design $D_m$ with 
$m=10$, for $\lambda = 1.227$.}

    \label{fig4results}
\end{figure}

\begin{figure}[ht!]
    \centering
    \includegraphics[width=0.55\textwidth]{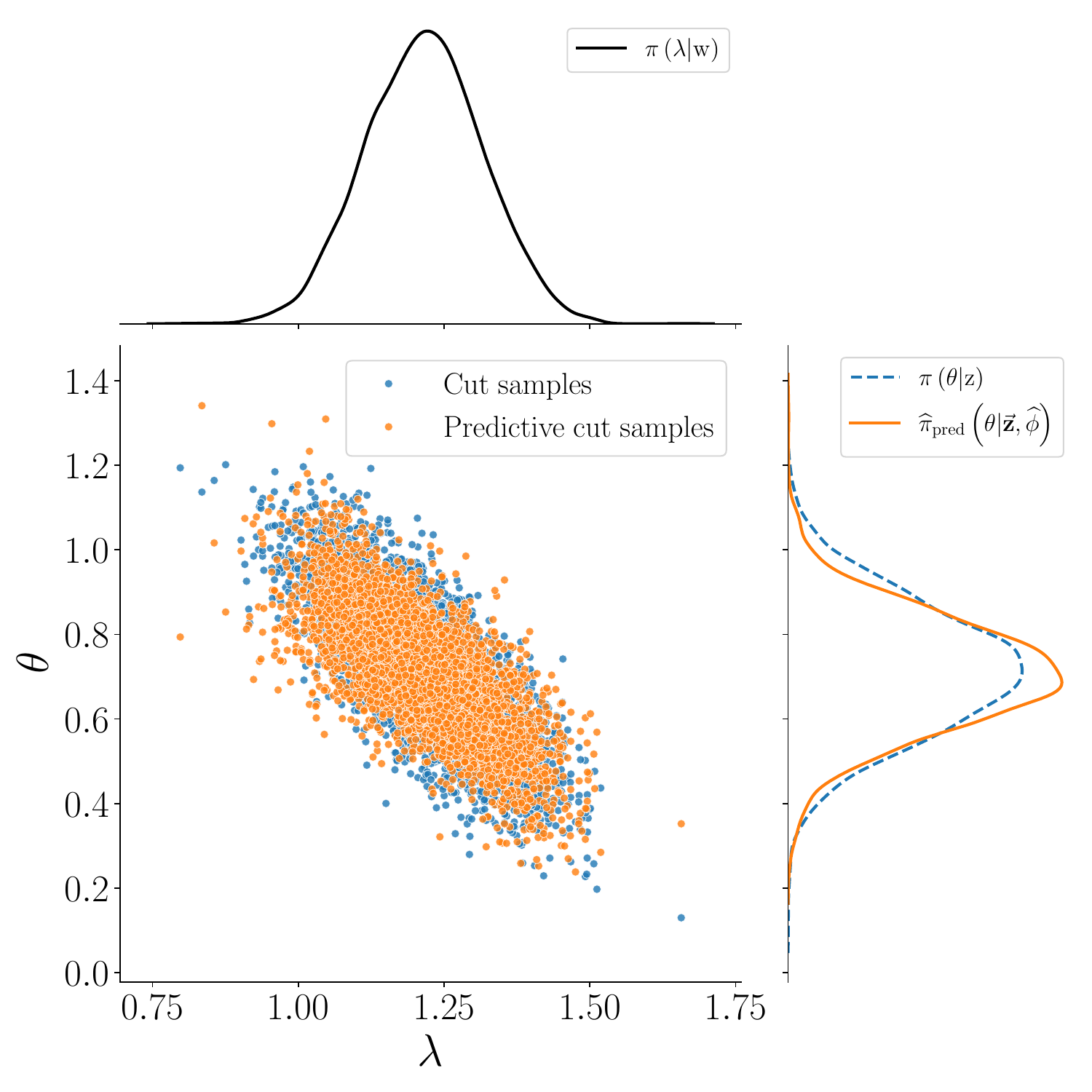}
\caption{Case $x_i \sim \mathcal{U}(-6,6)$ ($1 \leq i \leq n = 15$). Comparison between the target cut 
distribution under the Jeffreys prior on $(\lambda,\theta)$ and the GP--LinCC predictive cut distribution 
combined with $T^{-1}$, using a design $D_m$ of size $m=10$. For each distribution, $2000$ samples are displayed.}
    \label{fig5results}
\end{figure}

\begin{figure}
    \centering
    \includegraphics[width=0.6\textwidth]{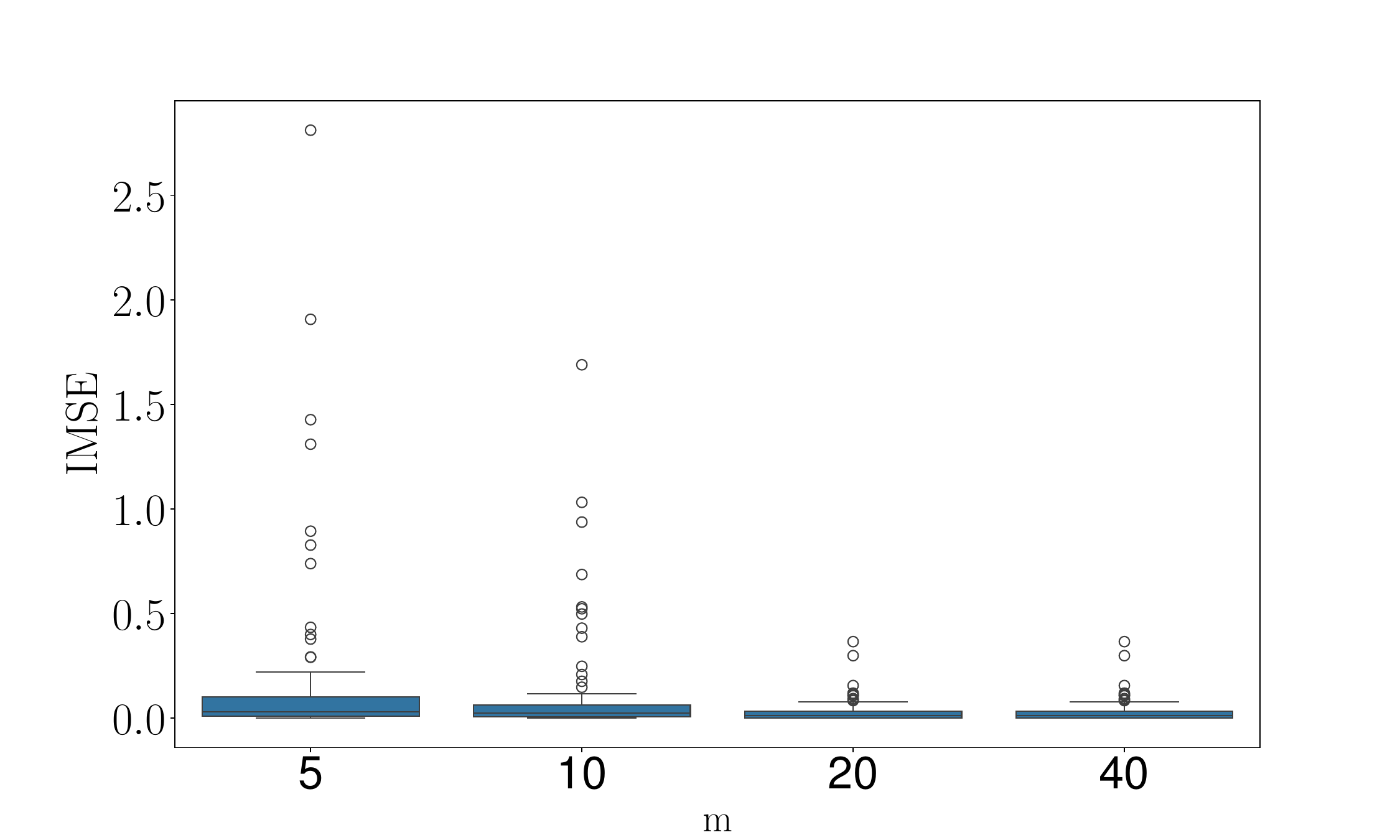}
\caption{Case $x_i \sim \mathcal{U}(-6,6)$ ($1 \leq i \leq n = 15$). Boxplot of the IMSE criterion for different sizes $m\in \{5, 10,20, 40 \}$ of $D_m$ and for a fixed sample $z$ of size $n=15$. Each boxplot is obtained from $1000$ samples.}

    \label{figimse2}
\end{figure}

\begin{figure}[ht!]
    \centering
    \includegraphics[width=0.5\textwidth]{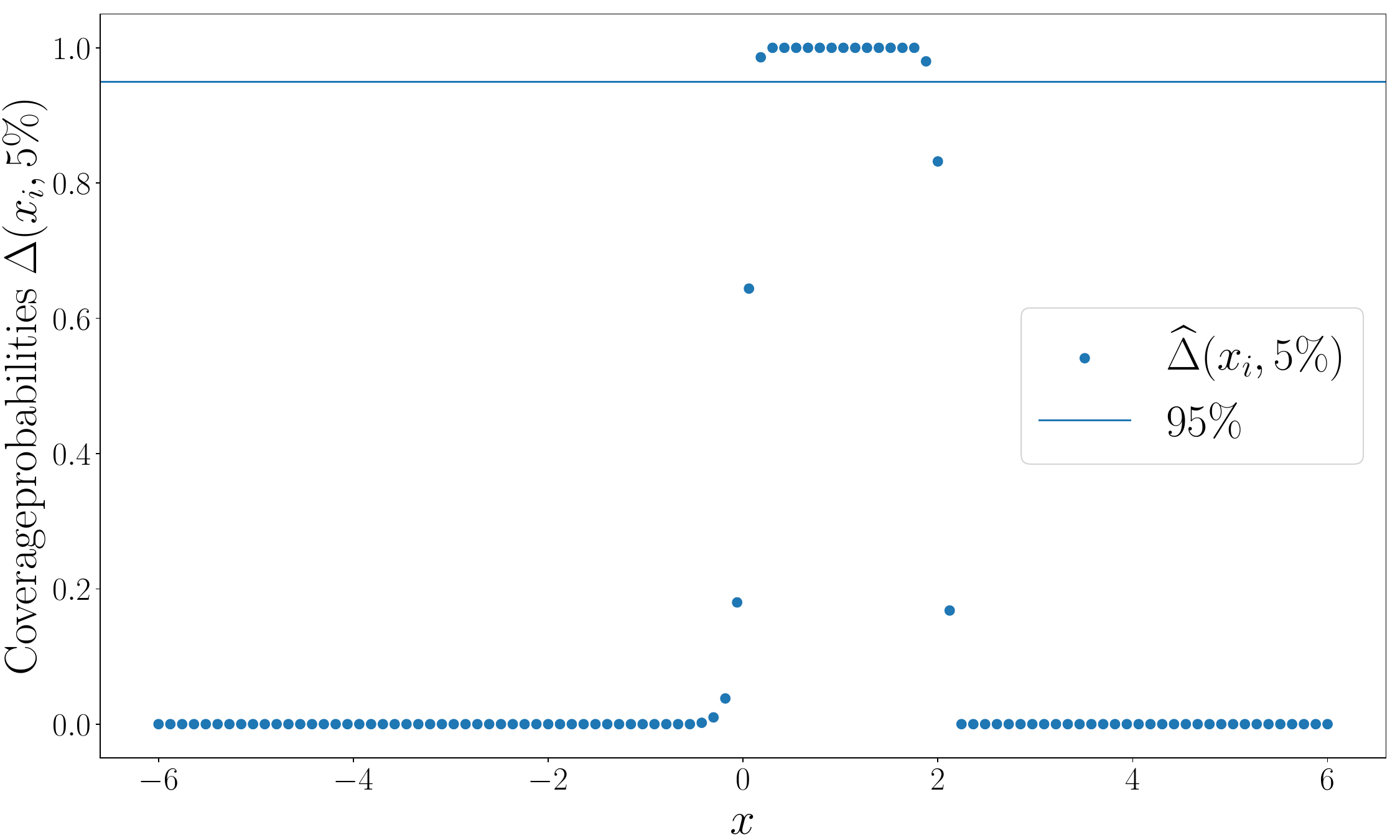}
    
    \caption{
    Empirical coverage probabilities $\hat{\Delta}(\alpha,x_i)$ 
    (Eq.~\eqref{eq53}, 100 sampled $x_i$). 
    }
    \label{fig7results}
\end{figure}

In this identifiable configuration, Figures~\ref{fig4results}, \ref{fig5results} and \ref{figimse2} show that GP--LinCC still provides an accurate approximation of the target cut
distribution. As expected, the compensation effect no longer arises when the model
is identifiable. This is reflected in the empirical coverage probabilities
$\hat{\Delta}(\alpha,x_i)$ displayed in Figure~\ref{fig7results}(a), which remain
below the $95\%$ threshold for most values of $x_i$.

For values of $x_i$ close to the sample mean $\bar{x}$, the coverage may
nevertheless exceed $95\%$. This does not indicate a compensation effect.
Indeed, when $x_i \approx \bar{x}$, the term $x_i(\lambda_1-\lambda_2)$
becomes numerically close to $-\bar{x}(\lambda_1-\lambda_2)$, which is the difference between the posterior means
$\mu_{\theta|\lambda_1}-\mu_{\theta|\lambda_2}$. Because the posterior dispersion of $\theta(\lambda)$ is small, the stochastic difference $\theta(\lambda_1)-\theta(\lambda_2)$ in Eq. (\ref{eq:draw_coverage}) remains close to the difference between these posterior means. Consequently, the two contributions in $Z_{ij}$ nearly cancel each other. As a result, $Z_{ij}$ concentrates near zero for such $x_i$, yielding local coverage rates above $95\%$ even though no compensation effect is present. 

\subsection{A two-dimensional example}
We consider the following example:
\begin{equation}
\left\{
\begin{array}{lll}
\displaystyle w_j = \lambda + \epsilon_{w_j} & ; & \epsilon_{w_j} \sim \mathcal{N}(0,\sigma_w^2), \\[2mm]
\displaystyle z_i = g_{\lambda,0}(x_i) + g_{\lambda,1}(x_i)\,\theta + \epsilon_{z_i} & ; & \epsilon_{z_i} \sim \mathcal{N}(0,\sigma_z^2),
\end{array}
\right.
\end{equation}
where
\begin{equation}
g_{\lambda,0}(x_i) = (\lambda + 1)\,\sin(20\lambda + 1), \qquad
g_{\lambda,1}(x_i) = (x_i + 1,\; x_i^{2} - 1),
\qquad
\theta =
\begin{pmatrix}
\theta_1 \\[1mm]
\theta_2
\end{pmatrix},
\end{equation}
and \(\sigma_z^{2} = 0.1\), \(\sigma_w^{2} = 0.2\), while \(x_i \sim \mathcal{U}[0,3]\) for
\(1 \leq i \leq n = 30\). The marginal cut posterior density
\(\pi(\lambda|w)\) is still Gaussian,
\begin{equation}
\pi(\lambda|w) \sim \mathcal{N}(\mu_{\lambda,\mathrm{cut}}, \sigma_{\lambda,\mathrm{cut}}^{2}),
\qquad
\mu_{\lambda,\mathrm{cut}} = \bar{w}, \quad
\sigma_{\lambda,\mathrm{cut}}^{2} = \frac{\sigma_w^{2}}{n_1},
\end{equation}
where \(n_1 = 30\) is the size of \(w\). GP--LinCC is implemented by assigning to each component of \(\theta(\lambda)\)
a Matérn \(5/2\) Gaussian process with constant mean, consistent with the
assumptions used throughout the paper.

\begin{figure}
    \centering
    \includegraphics[width=0.7\textwidth]{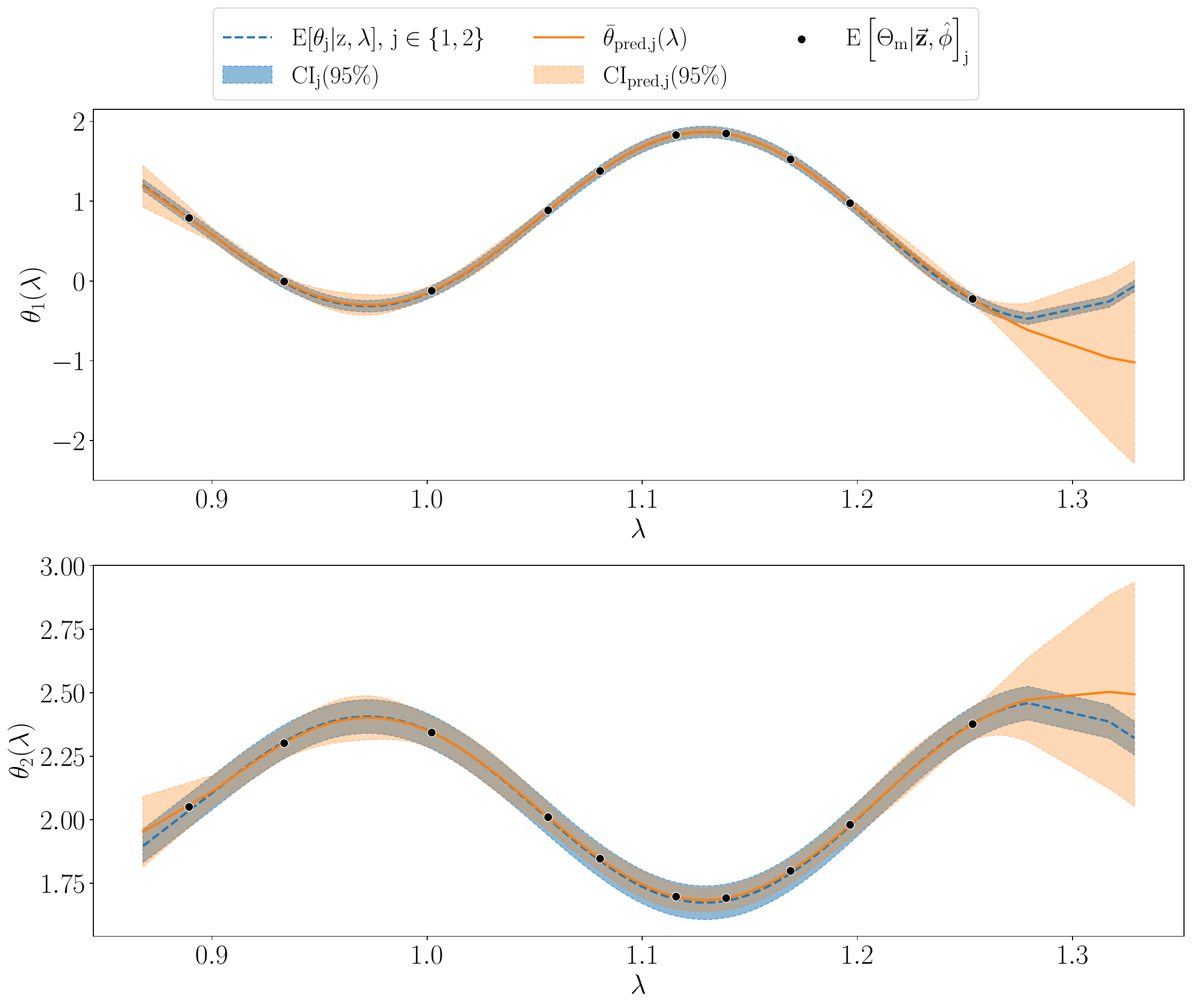}
    \caption{Comparison between the target conditional expectations 
$\mathbb{E}[\theta_j|\lambda,z]$ computed under $\pi(\theta|\lambda)\propto 1$ and their GP--LinCC predictive counterparts 
$\bar{\theta}_{\mathrm{pred},j}(\lambda)$, shown with $95\%$ credible intervals. 
The top panel displays $\theta_1$ and the bottom panel $\theta_2$ ($n=30$, $m=10$).}

    \label{example2Dfig1}
\end{figure}

\begin{figure}
    \centering
    \includegraphics[width=0.7\textwidth]{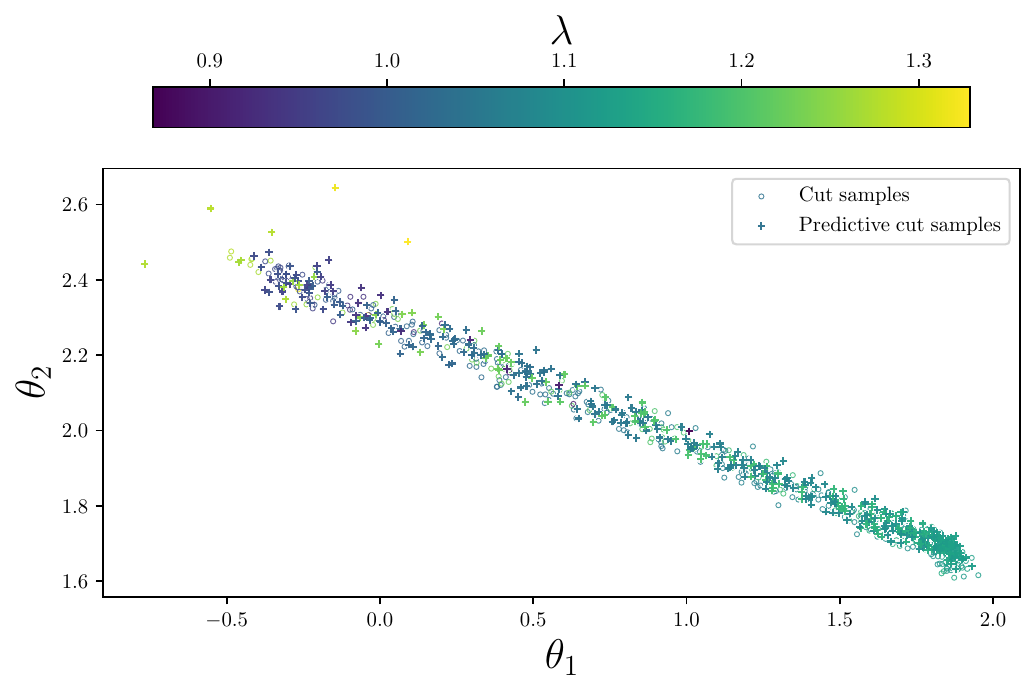}
\caption{Comparison between the target cut distribution $\pi_{\text{cut}}(\theta_1,\theta_2,\lambda|w,z)$ and the corresponding predictive ditribution $\pi_{\text{cut},\text{pred}}(\theta_1,\theta_2,\lambda|w,\vec{\zgris},\phi)$ computed by GP--LinCC.}

    \label{example2Dcut}
\end{figure}

\begin{figure}
    \centering
    \includegraphics[width=0.5\textwidth]{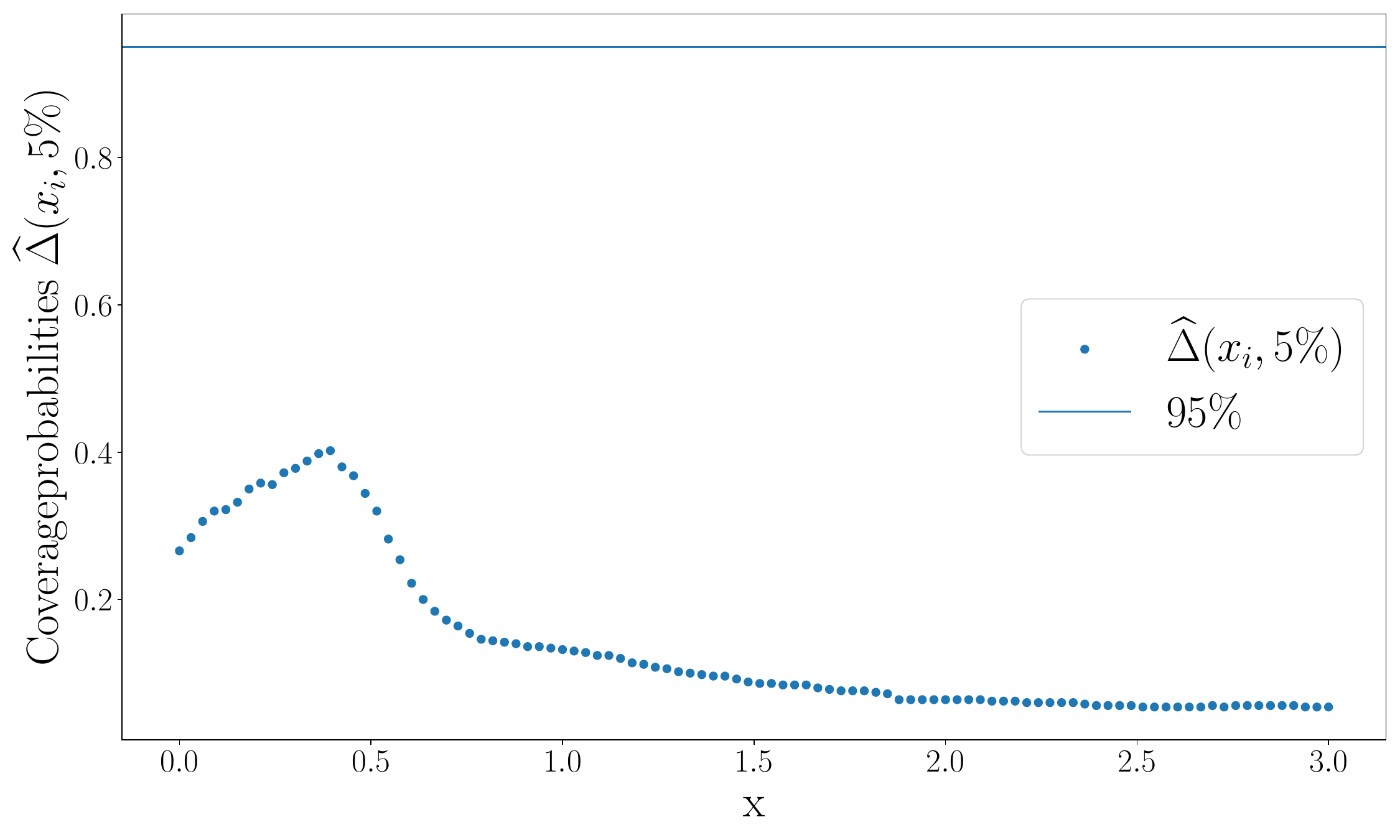}

\caption{(2-D example) Empirical coverage probabilities $\hat{\Delta}(\alpha,x_i)$ (Eq.~\eqref{eq34}, 
100 sampled $x_i$). 
}

    \label{example2Dfig2}
\end{figure}

\medskip
Figure~\ref{example2Dfig1} compares, as a function of \(\lambda\), the two
marginal densities extracted from \(\pi(\theta|\lambda,z)\) with the corresponding
predictive conditional densities obtained by GP--LinCC using \(n=30\) observations.
The predictive \(95\%\) predictive intervals for \(\theta_1(\lambda)\) and \(\theta_2(\lambda)\) cover
their target posterior means \(\mathbb{E}[\theta_1|\lambda,z]\) and
\(\mathbb{E}[\theta_2|\lambda,z]\) reasonably well, although the accuracy decreases
near the boundaries of the \(\lambda\)-domain. In fact, outside the range covered by the training samples, the GP--LinCC predictor operates in extrapolation, resulting in increasing deviations and widening predictive intervals.
Figure~\ref{example2Dcut} then compares the target and predictive cut
distributions, showing that the covariance structure of
\((\theta_1,\theta_2)\) as a function of \(\lambda\) is also well reproduced by
GP--LinCC. Finally, Figure~\ref{example2Dfig2} reports the empirical coverage probabilities
\(\widehat{\Delta}(5\%,x_i)\). As expected in this identifiable setting, the diagnostic confirms the absence of a
compensation effect.

\section{Conclusion}
\label{sec:conclusion}
In this paper, we have presented a new method to tackle the computation of a cut distribution within the framework of Bayesian calibration of two chained numerical models. Favoring such a distribution makes sense in multiphysics simulation where some experimental data may be unable to inform adequately part of the uncertain parameters of the numerical chain.
By assuming that the probability distribution of the parameter $\lambda$ is known, the method, called GP-LinCC, amounts to estimating the parameters $\theta$ of the downstream model conditionally on both the associated experimental data and the parameters $\lambda$ of the upstream model. This conditional distribution is derived from Bayes' theorem in which a Gaussian process prior is specified for the calibration function $\theta(\lambda)$ and the likelihood relies on a numerical design of the downstream model associated with a set $D_m$ of realizations for $\lambda$. As a result, GP-LinCC offers an analytical Gaussian posterior distribution provided that the downstream model is assumed to be a linear function of $\theta$, thus avoiding the use of an MCMC algorithm.
We further showed that GP-LinCC yields an analytically Gaussian predictive distribution for $\theta(\lambda^{\star})$ for any new realization $\lambda^{\star}\notin D_m$. 

GP-LinCC has been applied to some academic examples in low dimension for the parameter $\theta$, and the results obtained are convincing. Across these examples, the numerical experiments confirm that GP-LinCC provides accurate approximations of the exact conditional distribution $\pi(\theta|\lambda,z)$ involved in the cut formulation. The main discrepancies in fact occur near the extremes of the posterior support of $\lambda$, where the LHS design mapped through the Gaussian quantile function becomes quite sparse. These discrepancies can be mitigated by increasing the size of $D_m$, which overall reduces the gap between the exact and predictive cut distributions. However, the gap does not converge to $0$ because the GP-LinCC predictive distribution is structurally different from the exact conditional
$\pi(\theta|\lambda,z)$. The former relies on a Gaussian process approximation of $\theta(\lambda)$, combined with empirical-Bayes
hyperparameter estimation.


Several avenues for future work can be identified. First, the parameters $\theta$ may have bounded variation ranges related to their physical meaning. It would therefore be necessary to incorporate such bound constraints within the GP-LinCC framework to ensure that the method remains consistent when applied to real physical problems. In this case, the predictive distribution delivered by GP-LinCC becomes a truncated multivariate normal distribution \citep{da2020gaussian,lopez2018finite}. Another important aspect concerns the impact of linearization errors. In GP-LinCC, they are globally modelled through the scale parameters that capture the portion of variance unexplained by the linear emulators of the downstream model, but a more explicit assessment of their influence on the resulting cut distribution would be valuable. With highly time-consuming simulations, it would also be useful to develop adaptive designs for the set $D_m$ to improve the efficiency of the method. Finally, extending GP-LinCC to accommodate non-linear downstream models would further enhance its applicability and robustness for complex real-world problems.

We plan to apply the GP-LinCC method to the fuel application for pressurized water reactors that has motivated this methodological work, namely the calibration of the parameters of the fission gas behavior model conditionally on the thermal conductivity. However, a preliminary sensitivity analysis must be performed before deploying GP-LinCC on this physical problem. Indeed, the large dimension of $\theta$ (more than ten parameters) requires a pre-selection of the most influential parameters. To achieve this, we carried out a global sensitivity analysis using the multivariate version of the sensitivity indices based on the Hilbert--Schmidt independence criterion (HSIC) \citep{balde2025kernel}. 


\section*{Acknowledgments}
This work was partly funded under a tripartite project on Uncertainty Quantification between the French Alternative Energies and Atomic Energy Commission (CEA), Électricité de France (EDF), and Framatome (FRA). We thank Merlin Keller, research engineer at EDF R\&D, for insightful discussions on cut-off models, which contributed to the mathematical formalization of the calibration problem.

\section*{Declaration of generative AI and AI-assisted technologies in the manuscript preparation process}

During the preparation of this work, the authors used ChatGPT (OpenAI) in order to improve the English language and readability of the manuscript. After using this tool, the authors reviewed and edited the content as needed and take full responsibility for the content of the published article.

\bibstyle{abbrvnat}

\bibliography{references}

\newpage
\appendix

\section{Some useful mathematical results}
\numberwithin{equation}{subsection} 

\subsection{Vectorization}
\label{vec}

\noindent
Vectorization transforms any matrix \(A \in \mathbb{R}^{m \times p}\) into a
column vector \(\vec{A} \in \mathbb{R}^{mp}\) obtained by stacking the columns of
\(A\). For example,
\begin{equation}
A = 
\begin{pmatrix} a & b \\ c & d \end{pmatrix}
\qquad \Rightarrow \qquad
\vec{A} =
\begin{pmatrix} a \\ c \\ b \\ d \end{pmatrix}.
\end{equation}

\subsection{Gaussian process}
\label{gp}

A Gaussian process (GP) is a collection of random variables such that any finite
subset follows a multivariate normal distribution. Suppose that we observe 
\begin{equation}
\Theta_m = (\theta(\lambda_j))_{1 \leq j \leq m}^t \in \mathbb{R}^m.
\end{equation}
with \(\theta(\lambda_j)\in\mathbb{R}\), \(\lambda_j \in \mathbb{R}^q\) for some
\(q \ge 1\), and a design
\begin{equation}
D_m = (\lambda_1,\ldots,\lambda_m)^t.
\end{equation}
A GP prior is defined by its mean function \(m_\beta(\lambda)\) and its covariance
function
\[
C(\lambda,\lambda') := \sigma^2 K_\psi(\lambda,\lambda'),
\]
where \(K_\psi(\cdot,\cdot)\) is a correlation function depending on hyperparameters
\(\psi\), and \(\sigma^2>0\) is a variance parameter. 

\medskip
Let $\lambda^\star \in \R^{q}$ denote a new input location. The joint distribution of $\theta(\lambda^\star)$ and $\Theta_m$ is
\begin{equation}
\begin{pmatrix}
\theta(\lambda^\star) \\
\Theta_m
\end{pmatrix}
\sim
\mathcal{N}_{m+1}\!\left(
\begin{pmatrix}
m_\beta(\lambda^\star) \\
m_\beta(D_m)
\end{pmatrix},
\;
\begin{pmatrix}
C(\lambda^\star,\lambda^\star) & C(\lambda^\star, D_m) \\[1mm]
C(\lambda^\star, D_m)^t & C(D_m,D_m)
\end{pmatrix}
\right),
\end{equation}
where
\begin{equation}
m_\beta(D_m) = (m_\beta(\lambda_1),\ldots,m_\beta(\lambda_m))^t,
\end{equation}
\begin{equation}
C(D_m,D_m) = \big(C(\lambda_i,\lambda_j)\big)_{1\le i,j\le m},
\end{equation}
\begin{equation}
C(\lambda^\star, D_m)
=
\big(C(\lambda^\star,\lambda_j)\big)_{1\le j\le m},
\end{equation}
\begin{equation}
C(D_m,\lambda^\star) = C(\lambda^\star,D_m)^t.
\end{equation}
By the standard conditioning formulas for multivariate normal distributions,
assuming $C(D_m,D_m)$ is nonsingular, the GP predictive mean at $\lambda^\star$ is
\begin{equation}
\bar{\theta}(\lambda^\star)
=
m_\beta(\lambda^\star)
+
C(\lambda^\star,D_m)\,
C(D_m,D_m)^{-1}\,
\big(\Theta_m - m_\beta(D_m)\big),
\end{equation}
and the predictive covariance function is
\begin{equation}
\Sigma_{\mathrm{pred}}(\lambda^\star,\lambda^{\star'})
=
C(\lambda^\star,\lambda^{\star'})
-
C(\lambda^\star,D_m)\,
C(D_m,D_m)^{-1}\,
C(D_m,\lambda^{\star'}).
\end{equation}
The above predictive covariance corresponds to the standard
Gaussian process interpolation formulas assuming that the mean
parameters $\beta$ are fixed. If $\beta$ were estimated jointly
with the covariance hyperparameters, the predictive covariance
would contain an additional correction term, as in universal
kriging (see, e.g., \cite{rasmussen2006gaussian}).

The hyperparameters \(\phi := (\beta,\sigma^2,\psi)\), including regression and
covariance parameters, are unknown in practice and are typically estimated by
marginal likelihood maximization \citep{rasmussen2006gaussian}.

\medskip
In this appendix section, $\Theta_m$ is treated as observed to introduce the
standard Gaussian process interpolation formulas. In contrast, it is a latent quantity in the GP--LinCC framework.

\section{Analytical example of Section \ref{sec:bayesian}} 
\label{analytical}

\subsection{Expression of cut and full distributions}

We recall the statistical model,
\begin{equation}
\begin{pmatrix}
w \\
z
\end{pmatrix}
=
A_x
\begin{pmatrix}
\lambda \\
\theta
\end{pmatrix}
+
\begin{pmatrix}
\epsilon_w \\
\epsilon_z
\end{pmatrix}.
\end{equation}
with
\begin{equation}
A_x =
\begin{pmatrix}
\mathbf{1}_{n_1} & \mathbf{0}_{n_1} \\
x & \mathbf{1}_{n_2}
\end{pmatrix},
\qquad
x =(x_1,\cdots,x_{n_2})^t.
\end{equation}
and
\begin{equation}
\begin{pmatrix}
\epsilon_w \\
\epsilon_z
\end{pmatrix}
\sim
\mathcal{N}_{\,n_1+n_2}\!\left(
0,\;
\Sigma_{\sigma}:=
\begin{pmatrix}
\sigma^2_w I_{n_1} & 0 \\
0 & \sigma^2_z I_{n_2}
\end{pmatrix}
\right).
\end{equation}

\subsubsection*{Expression of the cut distribution \(\pi_{\mathrm{cut}}(\lambda, \theta|w, z)\)}

By Bayes formula, we have:
\begin{equation}
\pi(\theta|\lambda, z)
\propto
\mathcal{L}(z|\theta, \lambda)\,\pi(\theta),
\label{bayes_cut}
\end{equation}
where
\begin{equation}
\mathcal{L}(z|\theta, \lambda)
\propto
\prod_{i=1}^{n_2}
\exp\!\left(
-\frac{1}{2\sigma^2_z}\,
\big(z_i - (x_i\lambda + \theta)\big)^2
\right).
\end{equation}
After expansion of Eq.~(\ref{bayes_cut}), we obtain:
\begin{enumerate}

\item For Gaussian prior on \(\theta\):
\begin{equation}
\pi(\theta|\lambda, z)
\sim
\mathcal{N}(\mu_{\theta|\lambda, \mathrm{cut}}, \sigma^{2}_{\theta, \mathrm{cut}}).
\end{equation}
with
\begin{equation}
\sigma^{2}_{\theta, \mathrm{cut}}
=
\frac{\sigma^2_{\theta_0} (\sigma^2_z/n_2)}
{\sigma^2_{\theta_0}+ (\sigma^2_z/n_2)},
\qquad
\mu_{\theta|\lambda, \mathrm{cut}}
=
\frac{\sigma^2_{\theta_0}(\bar{z}- \bar{x}\lambda)
+ (\sigma^2_z/n_2)\theta_0}
{\sigma^2_{\theta_0}+ (\sigma^2_z/n_2)}.
\end{equation}

\item For Jeffreys prior on \(\theta\):
\begin{equation}
\pi(\theta|\lambda, z)
\sim
\mathcal{N}(\mu_{\theta|\lambda, \mathrm{cut}}, \sigma^{2}_{\theta, \mathrm{cut}}).
\end{equation}
with
\begin{equation}
\sigma^{2}_{\theta, \mathrm{cut}} = \frac{\sigma^2_z}{n_2},
\qquad
\mu_{\theta|\lambda, \mathrm{cut}} = \bar{z}-\bar{x}\lambda.
\end{equation}

\end{enumerate}

Note that, for Model \(1\) (i.e., \(\forall\,1\leq i\leq n_2, x_i = c\)), one has \(\bar{x}= c\).

Doing the same for \(\pi(\lambda|w)\), one has:

\begin{enumerate}

\item For Gaussian prior on \(\lambda\):
\begin{equation}
\pi(\lambda|w )
\sim
\mathcal{N}(\mu_{\lambda, \mathrm{cut}}, \sigma^2_{\lambda, \mathrm{cut}}).
\end{equation}
with
\begin{equation}
\mu_{\lambda, \mathrm{cut}}
=
\frac{\sigma^2_{\lambda_0} \bar{w}+ (\sigma^2_w/n_1)\lambda_0}
{\sigma^2_{\lambda_0}+ (\sigma^2_w/n_1)},
\qquad
\sigma^2_{\lambda, \mathrm{cut}}
=
\frac{\sigma^2_{\lambda_0} (\sigma^2_w/n_1)}
{\sigma^2_{\lambda_0}+ (\sigma^2_w/n_1)}.
\end{equation}

\item For Jeffreys prior on \(\lambda\):
\begin{equation}
\pi(\lambda|w)
\sim
\mathcal{N}(\mu_{\lambda, \mathrm{cut}}, \sigma^2_{\lambda, \mathrm{cut}}).
\end{equation}
with
\begin{equation}
\mu_{\lambda, \mathrm{cut}} = \bar{w},
\qquad
\sigma^2_{\lambda, \mathrm{cut}} = \frac{\sigma^2_w}{n_1}.
\end{equation}

\end{enumerate}
Finally, the expression of \(\pi_{\mathrm{cut}}(\theta, \lambda|w, z)\) is given by
\begin{equation}
\pi_{\mathrm{cut}}(\theta,\lambda|w, z)
\sim
\mathcal{N}(\mu_{\theta|\lambda, \mathrm{cut}}, \sigma^{2}_{\theta, \mathrm{cut}})
\otimes
\mathcal{N}(\mu_{\lambda, \mathrm{cut}}, \sigma^2_{\lambda, \mathrm{cut}}).
\end{equation}

\subsubsection*{Expression of the full distribution $\pi_{\mathrm{full}}(\lambda,\theta|w,z)$}

From Bayes' rule:
\begin{equation}
\pi_{\mathrm{full}}(\xi | w, z) \propto \lik(w, z | \xi)\,\pi(\xi),
\qquad
\xi = (\lambda, \theta)^t.
\end{equation}

\begin{enumerate}

\item \textbf{Gaussian prior on $\xi$}.  
The prior density is Gaussian:
\begin{equation}
\pi(\xi) \sim \mathcal{N}(\xi_0,\Sigma_0).
\end{equation}
Then
\begin{equation}
\pi_{\mathrm{full}}(\xi|w,z) \sim \mathcal{N}(\Sigma v, \Sigma)
\end{equation}
with
\begin{equation}
v = A_x^t \Sigma_{\sigma}^{-1} \begin{pmatrix}
w \\
z
\end{pmatrix} + \Sigma_0^{-1} \xi_0,
\qquad
\Sigma = \left(A_x^t \Sigma_{\sigma}^{-1} A_x + \Sigma_0^{-1}\right)^{-1}.
\end{equation}

\item \textbf{Jeffreys prior on $\xi$}.  
The prior density is constant:
\begin{equation}
\pi(\xi) \propto 1.
\end{equation}
Then
\begin{equation}
\pi_{\mathrm{full}}(\xi|w,z) \sim \mathcal{N}(\Sigma v, \Sigma)
\end{equation}
with
\begin{equation}
v = A_x^t \Sigma_{\sigma}^{-1} \begin{pmatrix}
w \\
z
\end{pmatrix},
\qquad
\Sigma = \left(A_x^t \Sigma_{\sigma}^{-1} A_x\right)^{-1}.
\end{equation}

We can then derive the marginal posterior of $\lambda$, which is also Gaussian:
\begin{equation}
\pi(\lambda | w, z)
=
\int \pi_{\mathrm{full}}(\lambda, \theta | w,z)\, d\theta
\sim \mathcal{N}\big((\Sigma v)_1,\; \Sigma_{11}\big).
\end{equation}
where $(\Sigma v)_1$ stands for the first coordinate of the vector $\Sigma v$, that is the posterior mean of $\lambda$, and $\Sigma_{11}$ denotes the first diagonal element of $\Sigma$, which is the posterior variance of $\lambda$.
\end{enumerate}

\subsection{KL divergence between $\pi_{\mathrm{cut}}(\lambda, \theta | w, z)$ and $\pi_{\mathrm{full}}(\lambda, \theta | w, z)$}
\label{B.1}

From Eq.~\eqref{eq6}, the KL divergence between $\pi_{\mathrm{cut}}(\lambda, \theta | w, z)$ and $\pi_{\mathrm{full}}(\lambda, \theta | w, z)$ reduces to the KL divergence between $\pi(\lambda | w, z)$ and $\pi(\lambda | w)$, which are Gaussian distributions. The KL divergence is therefore
\begin{equation}
\begin{aligned}
\mathrm{KL}\!\left(\pi(\lambda | w, z)\,\Vert\,\pi(\lambda | w)\right)
&=
\frac{1}{2}\,\log\!\left(\frac{\sigma^2_{\lambda,\mathrm{cut}}}{\Sigma_{11}}\right)
+
\frac{1}{2}\,
\frac{\Sigma_{11} + \big(\mu_{\lambda,\mathrm{cut}} - (\Sigma v)_1\big)^2}{\sigma^2_{\lambda,\mathrm{cut}}}
-
\frac{1}{2}.
\end{aligned}
\end{equation}
For Gaussian distributions, we know the following equivalence:
\begin{equation}
\mathrm{KL}\!\left(\pi(\lambda | w, z)\,\Vert\,\pi(\lambda | w)\right) = 0
\;\;\iff\;\;
(\Sigma v)_1 = \mu_{\lambda,\mathrm{cut}}
\quad\text{and}\quad
\Sigma_{11} = \sigma^2_{\lambda,\mathrm{cut}}.
\label{kl2}
\end{equation}
Let us see in which case Eq. \eqref{kl2} is satisfied. Based on the following expansion
\begin{equation}
    \begin{array}{ccc}
       A_x^t\Sigma_{\sigma}^{-1}A_x + \Sigma_0^{-1} &=& 
       \begin{pmatrix} \mathbf{1}^t_{n_1} & x^t \\\mathbf{0}^t_{n_1}  & \mathbf{1}^t_{n_2} \end{pmatrix}
       \begin{pmatrix} \sigma^{-2}_w I_{n_1} & \mathbf{0}_{n_1, n_2} \\ \mathbf{0}_{n_2, n_1} & \sigma^{-2}_z I_{n_2} \end{pmatrix}
       \begin{pmatrix} \mathbf{1}_{n_1} & \mathbf{0}_{n_1} \\ x & \mathbf{1}_{n_2} \end{pmatrix}  +  \Sigma_0^{-1} \\
       &\\  
       &=& \begin{pmatrix}
           n_1 \sigma^{-2}_w + \sigma^{-2}_z x^t x + \sigma^{-2}_{\lambda_0} & \sigma^{-2}_z x^t \mathbf{1}_{n_2} \\
          \\
            \sigma^{-2}_z\mathbf{1}^t_{n_2} x &  \sigma^{-2}_z n_2 +  \sigma^{-2}_{\theta_0}
       \end{pmatrix}
    \end{array}
\end{equation}
the matrix $\Sigma$ and the vector $v$ are equal respectively to
\begin{equation}
\Sigma=
          \frac{1}{\mathrm{det}\left(A_x^t\Sigma_{\sigma}^{-1}A_x + \Sigma_0^{-1}\right)}  \begin{pmatrix}
             \sigma^{-2}_z n_2 +  \sigma^{-2}_{\theta_0} & - \sigma^{-2}_z n_2 \bar{x} \\ 
             \\
             -\sigma^{-2}_z n_2 \bar{x} & n_1 \sigma^{-2}_w + \sigma^{-2}_z n_2\overline{x^2} + \sigma^{-2}_{\lambda_0}
         \end{pmatrix}
\end{equation}
and
\begin{equation}
v= \begin{pmatrix}
             \sigma^{-2}_w n_1 \bar{w} +   \sigma^{-2}_z x^t z +   \sigma^{-2}_{\lambda_0} \lambda_0 \\
             \\
              \sigma^{-2}_z n_2 \bar{z} +  \sigma^{-2}_{\theta_0} \theta_0 
         \end{pmatrix}.
\end{equation}
Then, one has: 
\begin{equation}
\left(\Sigma v\right)_1 =  \frac{ \left( \sigma^{-2}_z n_2 +  \sigma^{-2}_{\theta_0}  \right)  \left( \sigma^{-2}_w n_1 \bar{w} +   \sigma^{-2}_z x^t z +   \sigma^{-2}_{\lambda_0} \lambda_0\right) 
         - \sigma^{-2}_z n_2 \bar{x} \left( \sigma^{-2}_z \mathbf{1}^t_{n_2} z +  \sigma^{-2}_{\theta_0} \theta_0  \right)}{\mathrm{det}\left( A_x^t\Sigma_{\sigma}^{-1}A_x + \Sigma_0^{-1}\right)}
\end{equation}
and
\begin{equation}
\Sigma_{11}= \frac{1}{\mathrm{det}\left(A_x^t\Sigma_{\sigma}^{-1}A_x + \Sigma_0^{-1}\right)} (\sigma^{-2}_z n_2 +  \sigma^{-2}_{\theta_0}). 
\end{equation}
with
\begin{multline}
    \mathrm{det}\left(A_x^t\Sigma_{\sigma}^{-1}A_x + \Sigma_0^{-1}\right) = \Big[
 (\sigma^{-2}_z n_2 +  \sigma^{-2}_{\theta_0} )\times \\
 (n_1 \sigma^{-2}_w + \sigma^{-2}_z n_2 \overline{x^2} + \sigma^{-2}_{\lambda_0})\Big] - (\sigma^{-2}_z n_2 \bar{x})^2.
\end{multline}
It follows that
\begin{align}
\nonumber
\Sigma_{11} = \sigma^2_{\lambda, \mathrm{cut}}
&\iff
\label{eq:pre_jensen}
\nonumber
\\
&\frac{\sigma^{-2}_z n_2 + \sigma^{-2}_{\theta_0}}
{
(\sigma^{-2}_z n_2 + \sigma^{-2}_{\theta_0})
\left(
n_1 \sigma^{-2}_w
+ \sigma^{-2}_z n_2 \overline{x^2}
+ \sigma^{-2}_{\lambda_0}
\right)
-
\left(\sigma^{-2}_z n_2 \bar{x}\right)^2
}
\nonumber
\\
&\qquad =
\frac{\sigma^2_{\lambda_0}(\sigma^2_w/n_1)}
{\sigma^2_{\lambda_0}+(\sigma^2_w/n_1)}
\nonumber
\\[2mm]
&\iff
(\sigma^{-2}_z n_2 + \sigma^{-2}_{\theta_0})
\sigma^{-2}_z n_2 \overline{x^2}
=
(\sigma^{-2}_z n_2 \bar{x})^2
\nonumber
\\[2mm]
&\iff
(\sigma^{-2}_z n_2 + \sigma^{-2}_{\theta_0})\overline{x^2}
=
\sigma^{-2}_z n_2 \bar{x}^2 .
\end{align}
Note that: 
\begin{equation}
    \begin{array}{cccc}
         \sigma^{-2}_z n_2 +  \sigma^{-2}_{\theta_0} & >& \sigma^{-2}_z n_2, \\ 
          \overline{x^2} &\geq& \overline{x}^2  &\text{(Jensen inequality in the discrete case)}.
    \end{array}
\end{equation}
Combining Eq.~(\ref{eq:pre_jensen}) with the above inequalities implies that
\begin{equation}
\overline{x^2}=0
\qquad\text{and}\qquad
\overline{x}^2=0.
\end{equation}
Since $\overline{x^2} = \frac{1}{n_2}\sum_{i=1}^{n_2}x_i^2$, this implies that
$x_i=0$ for all $1\le i\le n_2$. Therefore,
\begin{equation}
\left(\sigma^{-2}_z n_2 + \sigma^{-2}_{\theta_0}\right)\overline{x^2}
=
\sigma^{-2}_z n_2 \overline{x}^2
\iff
\forall\,1\le i\le n_2,\; x_i=0 .
\end{equation}
A similar computation shows that the condition 
$(\Sigma v)_1 = \mu_{\lambda,\mathrm{cut}}$ does not, by itself, impose
$x_i = 0$ for all $i$. However, once $x_i = 0$ is enforced by the
variance condition, one verifies that $(\Sigma v)_1 = \mu_{\lambda,\mathrm{cut}}$
automatically holds. Hence both conditions in Eq.~\eqref{kl2} are
simultaneously satisfied if and only if $x_i=0$ for all $1\le i\le n_2$.
Therefore, under a Gaussian prior, we have established the following result:
\begin{equation}
\mathrm{KL}\!\left(\pi(\lambda|w,z)\,\Vert\,\pi(\lambda|w)\right)=0
\iff
\forall\,1\le i \le n_2,\; x_i = 0.
\end{equation}

\medskip
For Jeffreys prior, following exactly the same reasoning as in the Gaussian–prior
case, we obtain
\begin{align}
\nonumber
         \Sigma_{11} = \sigma^2_{\lambda,\mathrm{cut}}
         \iff&
         \;\left(\sigma^{-2}_z n_2\right)\,\overline{x^2}
         =
         \sigma^{-2}_z n_2\,\overline{x}^2 \\[1mm]
\nonumber
    \iff&\;\overline{x^2} = \overline{x}^2 \\[1mm]
\iff&\;\forall\,1\le i\le n_2,\; x_i = c\in\mathbb{R}.
\end{align}
As in the Gaussian prior case, the condition
$(\Sigma v)_1 = \mu_{\lambda,\mathrm{cut}}$ does not, by itself, force
$x_i = c$ for all $i$. However, once $x_i = c$ is imposed by the variance
condition, one checks that $(\Sigma v)_1 = \mu_{\lambda,\mathrm{cut}}$
automatically holds. Therefore, both conditions in Eq.~\eqref{kl2} are
satisfied if and only if $x_i = c$ for all $1 \le i \le n_2$. Thus, we have
\begin{equation}
         \mathrm{KL}\!\left(\pi(\lambda|w,z)\,\Vert\,\pi(\lambda|w)\right)=0
         \iff
         \forall\,1\le i\le n_2,\; x_i = c \in \mathbb{R}.
\end{equation}

\section{Proof of the results of Section \ref{sec:gplincc}}

\subsection{Proof of Theorem \ref{theo1}}
\label{B.1}

\begin{proof}
In the GP--LinCC framework, the vectorized parameter
$\vec{\Theta}_m$ stacks the $p$ model parameters evaluated
at the $m$ design points, so that $\vec{\Theta}_m\in\mathbb{R}^{pm}$.
Consequently, the covariance matrix $\C_\phi$ is of dimension
$pm\times pm$.

We consider the vectorized model
\begin{equation}
    \vec{\mathbf{z}}
    =
    G\,\vec{\Theta}_m + \vec{\epsilon},
    \qquad
    \vec{\epsilon} \sim \mathcal{N}(0,\Sigma_{\vec{\epsilon}}),
\end{equation}
together with the Gaussian prior
\begin{equation}
    \vec{\Theta}_m \sim \mathcal{N}(\vec{M}_\beta, \C_\phi).
\end{equation}
We assume that \(\C_\phi\) is symmetric positive definite, so that \(\C_\phi^{-1}\) exists. 
Since \(\vec{\epsilon}\) is independent of \(\vec{\Theta}_m\), the joint vector
\begin{equation}
\begin{pmatrix}
\vec{\Theta}_m \\
\vec{\mathbf{z}}
\end{pmatrix}
\end{equation}
is multivariate normal with mean
\begin{equation}
\begin{pmatrix}
\vec{M}_\beta \\
G\vec{M}_\beta
\end{pmatrix}
\end{equation}
and block covariance matrix
\begin{equation}
\begin{pmatrix}
\C_\phi & \C_\phi G^t \\
G\C_\phi & \Sigma_{\vec{\epsilon}} + G\C_\phi G^t
\end{pmatrix}.
\end{equation} 
The posterior distribution is therefore Gaussian, and its mean and covariance are obtained from the standard conditioning formulas for multivariate normal vectors:
\begin{equation}
\begin{aligned}
    \Sigma_\phi &= \left(\Delta^{-1} + \C_\phi^{-1}\right)^{-1}, \\[2mm]
    \mu_\phi &= \Sigma_\phi\left(G^t\Sigma_{\vec{\epsilon}}^{-1}\vec{\mathbf{z}} + \C_\phi^{-1}\vec{M}_\beta\right).
\end{aligned}
\label{eq:posterior_generic}
\end{equation}
with $\Delta^{-1}:=G^t\Sigma_{\vec{\epsilon}}^{-1}G$.
\end{proof}

\subsection{Proof of Theorem \ref{theo2}}
\label{B.2}
\begin{proof}
The predictive distribution of \(\theta(\lambda^\star)\), for a fixed $\lambda^\star$, is defined by
\begin{equation}
\label{eq:int_pred_bis}
\pi_{\mathrm{pred}}\big(\vec{\theta}(\lambda^\star)| \vec{\mathbf{z}},\phi\big)
=
\int_{\mathcal{T}^m}
\pi\big(\vec{\theta}(\lambda^\star)| \vec{\Theta}_m,\phi\big)\,
\pi\big(\vec{\Theta}_m| \vec{\mathbf{z}},\phi\big)\,
d\vec{\Theta}_m.
\end{equation}
Set \(\vec{\theta}^\star := \vec{\theta}(\lambda^\star)\). From the GP prior specification, the joint vector
\begin{equation}
\begin{pmatrix}
\vec{\Theta}_m \\
\vec{\theta}^\star
\end{pmatrix}
\end{equation}
is multivariate normal with mean
\begin{equation}
\begin{pmatrix}
\vec{M}_\beta \\
\vec{m}_\beta(\lambda^\star)
\end{pmatrix}
\end{equation}
and block covariance matrix
\begin{equation}
\begin{pmatrix}
\C_\phi & \C(D_m,\lambda^\star) \\
\C(\lambda^\star,D_m) & \C(\lambda^\star,\lambda^\star)
\end{pmatrix}.
\end{equation}
By the standard conditioning formulas for multivariate normal distributions (see, e.g., Appendix~A.2 in \cite{rasmussen2006gaussian}), the conditional distribution of \(\theta^\star\) given \(\vec{\Theta}_m\) and \(\phi\) is
\begin{equation}
\label{eq:cond_theta_given_Theta}
\vec{\theta}^\star | \vec{\Theta}_m,\phi
\sim
\mathcal{N}\!\Big(
\vec{m}_\beta(\lambda^\star)
+
\C(\lambda^\star,D_m)\C_\phi^{-1}
\big(\vec{\Theta}_m-\vec{M}_\beta\big),
\;
\Sigma_{\mathrm{cond}}(\lambda^\star,\lambda^\star)
\Big),
\end{equation}
with
\begin{equation}
\Sigma_{\mathrm{cond}}(\lambda^\star,\lambda^\star)
=
\C(\lambda^\star,\lambda^\star)
-
\C(\lambda^\star,D_m)\C_\phi^{-1}
\C(D_m,\lambda^\star).
\end{equation}
Equivalently,
\begin{equation}
\label{eq:theta_star_linear_repr}
\vec{\theta}^\star
=
\vec{m}_\beta(\lambda^\star)
+
\C(\lambda^\star,D_m)\C_\phi^{-1}
\big(\vec{\Theta}_m-\vec{M}_\beta\big)
+
\eta_{\mathrm{cond}},
\end{equation}
where \(\eta_{\mathrm{cond}}\sim\mathcal{N}\!\big(0,\Sigma_{\mathrm{cond}}(\lambda^\star,\lambda^\star)\big)\) is independent of \(\vec{\Theta}_m\). Since \(\pi(\vec{\Theta}_m| \vec{\mathbf{z}},\phi)\) is Gaussian, integrating \(\pi(\vec{\theta}^\star| \vec{\Theta}_m,\phi)\) with respect to this posterior yields a Gaussian distribution. By Theorem~\ref{theo1},
\begin{equation}
\vec{\Theta}_m | \vec{\mathbf{z}},\phi
\sim
\mathcal{N}(\mu_\phi,\Sigma_\phi).
\end{equation}
Substituting \(\vec{\Theta}_m = \mu_\phi + (\vec{\Theta}_m-\mu_\phi)\) into Eq.~\eqref{eq:theta_star_linear_repr} gives
\begin{multline}
\label{eq:theta_star_given_z_phi}
\vec{\theta}^\star | \vec{\mathbf{z}},\phi
=
\vec{m}_\beta(\lambda^\star)
+
\C(\lambda^\star,D_m)\C_\phi^{-1}
\big(\mu_\phi-\vec{M}_\beta\big)
+\\
\C(\lambda^\star,D_m)\C_\phi^{-1}
\big(\vec{\Theta}_m-\mu_\phi\big)
+
\eta_{\mathrm{cond}}.
\end{multline}
Therefore \(\vec{\theta}^\star| \vec{\mathbf{z}},\phi\) is Gaussian with mean
\begin{equation}
\bar{\theta}_{\mathrm{pred}}(\lambda^\star)
=
\vec{m}_\beta(\lambda^\star)
+
\C(\lambda^\star,D_m)\C_\phi^{-1}
\big(\mu_\phi-\vec{M}_\beta\big),
\end{equation}
which coincides with Eq.~\eqref{eq:pred_mean}. Moreover,
\begin{equation}
\vec{\theta}^\star - \bar{\theta}_{\mathrm{pred}}(\lambda^\star)
=
\C(\lambda^\star,D_m)\C_\phi^{-1}
\big(\vec{\Theta}_m-\mu_\phi\big)
+
\eta_{\mathrm{cond}}.
\end{equation}
Using independence and zero means,
\begin{equation}
\Sigma_{\mathrm{pred}}(\lambda^\star,\lambda^\star)
=
\C(\lambda^\star,D_m)\C_\phi^{-1}
\Sigma_\phi
\C_\phi^{-1}
\C(D_m,\lambda^\star)
+
\Sigma_{\mathrm{cond}}(\lambda^\star,\lambda^\star).
\end{equation}
For a general pair \((\lambda^\star,\lambda^{\star\prime})\), define
\begin{equation}
\Sigma_{\mathrm{cond}}(\lambda^\star,\lambda^{\star\prime})
=
\C(\lambda^\star,\lambda^{\star\prime})
-
\C(\lambda^\star,D_m)\C_\phi^{-1}
\C(D_m,\lambda^{\star\prime}).
\end{equation}
Then the predictive cross-covariance is given by
\begin{equation}
\Sigma_{\mathrm{pred}}(\lambda^\star,\lambda^{\star\prime})
=
\Sigma_{\mathrm{cond}}(\lambda^\star,\lambda^{\star\prime})
+
\C(\lambda^\star,D_m)\C_\phi^{-1}
\Sigma_\phi
\C_\phi^{-1}
\C(D_m,\lambda^{\star\prime}).
\end{equation}
which is Eq.~\eqref{eq:pred_cov}. This completes the proof.
\end{proof}

\subsection{Marginal likelihood expression and hyperparameters tuning}
\label{B.4}

\noindent
The hyperparameters \(\phi=\{(\beta_l,\sigma_l^2,\psi_l)\}_{l=1}^p\) are estimated by maximizing the marginal likelihood of the vectorized experimental data \(\vec{\mathbf{z}} \in \mathbb{R}^{nm}\). Conditionally on \(\vec{\Theta}_m\), the likelihood is given by
\begin{equation}
\lik(\vec{\mathbf{z}}|\vec{\Theta}_m)\sim\mathcal{N}\!\left(\vec{\mathbf{z}};\, G\vec{\Theta}_m,\Sigma_{\vec{\epsilon}}\right),
\end{equation}
while the prior distribution of \(\vec{\Theta}_m\) reads
\begin{equation}
\pi(\vec{\Theta}_m|\phi)\sim\mathcal{N}\!\left(\vec{\Theta}_m;\,\vec{M}_\beta,\C_\phi\right),
\end{equation}
where \(\C_\phi\) is assumed to be symmetric positive definite. The marginal likelihood is obtained by integrating out \(\vec{\Theta}_m\):
\begin{equation}
\widehat{\phi}=\argmax_{\phi}\int_{\mathcal{T}^m}\lik(\vec{\mathbf{z}}|\vec{\Theta}_m)\,\pi(\vec{\Theta}_m|\phi)\,d\vec{\Theta}_m .
\end{equation}

\noindent
Since the likelihood and the prior define a jointly Gaussian model in \((\vec{\mathbf{z}},\vec{\Theta}_m)\), the marginal distribution of \(\vec{\mathbf{z}}\) is Gaussian and is obtained from standard linear--Gaussian marginalization formulas:
\begin{equation}
\pi(\vec{\mathbf{z}}|\phi)\sim\mathcal{N}\!\left(\vec{\mathbf{z}};\,G\vec{M}_\beta,\,\Sigma_{\vec{\epsilon}}+G\C_\phi G^t\right).
\end{equation}
Define for convenience the marginal covariance matrix
\begin{equation}
\Sigma_{\mathbf{z}}(\phi):=\Sigma_{\vec{\epsilon}}+G\C_\phi G^t\in\mathbb{R}^{nm\times nm}.
\end{equation}
The marginal likelihood can therefore be written explicitly as
\begin{equation}
\pi(\vec{\mathbf{z}}|\phi)=\frac{\exp\!\left(-\tfrac12(\vec{\mathbf{z}}-G\vec{M}_\beta)^t\Sigma_{\mathbf{z}}(\phi)^{-1}(\vec{\mathbf{z}}-G\vec{M}_\beta)\right)}{(2\pi)^{nm/2}\big|\Sigma_{\mathbf{z}}(\phi)\big|^{1/2}}.
\end{equation}
Taking logarithms yields
\begin{equation}
\log \pi(\vec{\mathbf{z}}|\phi)=-\tfrac12\log\!\left|\Sigma_{\mathbf{z}}(\phi)\right|-\tfrac12(\vec{\mathbf{z}}-G\vec{M}_\beta)^t\Sigma_{\mathbf{z}}(\phi)^{-1}(\vec{\mathbf{z}}-G\vec{M}_\beta)+\mathrm{const}.
\end{equation}
For convenience, we introduce the negative marginal log-likelihood
\begin{equation}
\Bell(\beta,\sigma^2,\psi):=-\,2\log \pi(\vec{\mathbf{z}}|\phi),
\end{equation}
where the additive constant independent of \(\phi\) is omitted. Assume now that the prior mean has the linear form
\begin{equation}
\vec{M}_\beta=H\beta,\qquad H\in\mathbb{R}^{pm\times q},\;\;\beta\in\mathbb{R}^{q},
\end{equation}
with \(H\) of full column rank. Then the criterion becomes
\begin{equation}
\Bell(\beta,\sigma^2,\psi)=\log\!\left|\Sigma_{\mathbf{z}}(\phi)\right|+(\vec{\mathbf{z}}-GH\beta)^t\Sigma_{\mathbf{z}}(\phi)^{-1}(\vec{\mathbf{z}}-GH\beta)+\mathrm{const},
\end{equation}
which shows that, for fixed \((\sigma^2,\psi)\), the dependence of \(\Bell\) on \(\beta\) is entirely contained in a generalized least-squares quadratic form.
\begin{proposition}\label{prop4}
Assume that \(\vec{M}_\beta = H\beta\) with \(H\in\mathbb{R}^{pm\times q}\) of full column rank, and define
\[
\Sigma_{\mathbf{z}}(\phi) := \Sigma_{\vec{\epsilon}} + G\C_\phi G^t.
\]
Then, for fixed \((\sigma^2,\psi)\), the minimizer of \(\Bell(\beta,\sigma^2,\psi)\) with respect to \(\beta\) is the generalized least-squares estimator
\begin{equation}
\label{eq:gls_estimator}
\hat{\beta}(\sigma^2,\psi)
=
\big(H^t G^t \Sigma_{\mathbf{z}}(\phi)^{-1} G H\big)^{-1}
H^t G^t \Sigma_{\mathbf{z}}(\phi)^{-1}\vec{\mathbf{z}}.
\end{equation}
\end{proposition}

\subsubsection*{Proof of Proposition \ref{prop4}}
\begin{proof}
For fixed \((\sigma^2,\psi)\), the marginal criterion
\[
\Bell(\beta,\sigma^2,\psi)
=
\log|\Sigma_{\mathbf{z}}(\phi)|
+
(\vec{\mathbf{z}}-GH\beta)^t
\Sigma_{\mathbf{z}}(\phi)^{-1}
(\vec{\mathbf{z}}-GH\beta)
+
\mathrm{const}
\]
depends on \(\beta\) only through the quadratic term. Minimizing \(\Bell\) with respect to \(\beta\) is therefore equivalent to minimizing
\begin{equation}
Q(\beta)
:=
(\vec{\mathbf{z}}-GH\beta)^t
\Sigma_{\mathbf{z}}(\phi)^{-1}
(\vec{\mathbf{z}}-GH\beta).
\end{equation}
Since \(\Sigma_{\mathbf{z}}(\phi)\) is symmetric positive definite, \(Q(\beta)\) is a strictly convex quadratic form in \(\beta\).
Taking the gradient with respect to \(\beta\) yields
\begin{equation}
\nabla_\beta Q(\beta)
=
-2\,H^t G^t \Sigma_{\mathbf{z}}(\phi)^{-1}
(\vec{\mathbf{z}}-GH\beta).
\end{equation}
The first-order optimality condition \(\nabla_\beta Q(\beta)=0\) gives
\begin{equation}
H^t G^t \Sigma_{\mathbf{z}}(\phi)^{-1} G H\,\beta
=
H^t G^t \Sigma_{\mathbf{z}}(\phi)^{-1}\vec{\mathbf{z}}.
\end{equation}
Since \(H\) has full column rank and \(\Sigma_{\mathbf{z}}(\phi)\) is positive definite, \(H^t G^t \Sigma_{\mathbf{z}}(\phi)^{-1} G H\) is invertible. Hence the unique minimizer is
\begin{equation}
\hat{\beta}(\sigma^2,\psi)
=
\big(H^t G^t \Sigma_{\mathbf{z}}(\phi)^{-1} G H\big)^{-1}
H^t G^t \Sigma_{\mathbf{z}}(\phi)^{-1}\vec{\mathbf{z}},
\end{equation}
which concludes the proof.
\end{proof}

\medskip
Based on Proposition~\ref{prop4}, the estimation of the hyperparameters
\((\sigma^2,\psi)\) reduces to the optimization of the marginal criterion
with the estimator \(\hat{\beta}(\sigma^2,\psi)\) plugged in:
\begin{equation}
\label{B.3.3}
(\widehat{\sigma}^2,\widehat{\psi})
=
\argmin_{\sigma^2,\psi}\;
\Bell\big(\hat{\beta}(\sigma^2,\psi),\sigma^2,\psi\big).
\end{equation}
The resulting profile objective function is
\begin{multline}
\label{eq:Bell-Theta-space}
\Bell\big(\hat{\beta}(\sigma^2,\psi),\sigma^2,\psi\big)
=
\log\!\left|\Sigma_{\vec{\epsilon}} + G \C_\phi G^t\right|
\\
+
\big(\vec{\mathbf{z}} - G\vec{M}_{\hat{\beta}}\big)^t
\big(\Sigma_{\vec{\epsilon}} + G \C_\phi G^t\big)^{-1}
\big(\vec{\mathbf{z}} - G\vec{M}_{\hat{\beta}}\big)
\end{multline}
where \(\vec{M}_{\hat{\beta}} := H\hat{\beta}(\sigma^2,\psi)\).
The minimization with respect to \((\sigma^2,\psi)\) is performed numerically.

\begin{remark}[Alternative expression of $\hat{\beta}$]
\label{rem:beta-woodbury}
Assume in addition that the design matrix $G$ has full column rank, so that
$\Delta^{-1}=G^t\Sigma_{\vec{\epsilon}}^{-1}G$ is invertible and $\Delta$ is well-defined.
Under the assumptions of Proposition~\ref{prop4}, the estimator
\begin{equation}
\hat{\beta}(\sigma^2,\psi)
=
\big(H^t G^t \Sigma_{\mathbf{z}}(\phi)^{-1} G H\big)^{-1}
H^t G^t \Sigma_{\mathbf{z}}(\phi)^{-1}\vec{\mathbf{z}},
\end{equation}
with
\begin{equation}
\Sigma_{\mathbf{z}}(\phi)=\Sigma_{\vec{\epsilon}}+G\C_\phi G^t,
\end{equation}
admits the equivalent representation
\begin{equation}
\label{eq:beta-woodbury}
\hat{\beta}(\sigma^2,\psi)
=
\big(H^t(\Delta+\C_\phi)^{-1}H\big)^{-1}
H^t(\Delta+\C_\phi)^{-1}
\Delta\,G^t\Sigma_{\vec{\epsilon}}^{-1}\vec{\mathbf{z}},
\end{equation}
where
\begin{equation}
\Delta^{-1}=G^t\Sigma_{\vec{\epsilon}}^{-1}G.
\end{equation}
\end{remark}

\begin{proof}
By Woodbury’s identity applied to
\begin{equation}
\Sigma_{\mathbf{z}}(\phi)=\Sigma_{\vec{\epsilon}}+G\C_\phi G^t,
\end{equation}
we obtain
\begin{equation}
\label{eq:woodbury-beta}
\Sigma_{\mathbf{z}}(\phi)^{-1}
=
\Sigma_{\vec{\epsilon}}^{-1}
-
\Sigma_{\vec{\epsilon}}^{-1}
G\big(\C_\phi^{-1}+G^t\Sigma_{\vec{\epsilon}}^{-1}G\big)^{-1}
G^t\Sigma_{\vec{\epsilon}}^{-1}.
\end{equation}
Using the definition
\begin{equation}
\Delta^{-1}=G^t\Sigma_{\vec{\epsilon}}^{-1}G,
\end{equation}
this expression can be rewritten as
\begin{equation}
\Sigma_{\mathbf{z}}(\phi)^{-1}
=
\Sigma_{\vec{\epsilon}}^{-1}
-
\Sigma_{\vec{\epsilon}}^{-1}
G(\C_\phi^{-1}+\Delta^{-1})^{-1}
G^t\Sigma_{\vec{\epsilon}}^{-1}.
\end{equation}
Substituting this expression into \(G^t\Sigma_{\mathbf{z}}(\phi)^{-1}\) yields
\begin{equation}
\label{eq:GtSigmazinv_step1}
G^t\Sigma_{\mathbf{z}}(\phi)^{-1}
=
\Big(I-\Delta^{-1}(\C_\phi^{-1}+\Delta^{-1})^{-1}\Big)
G^t\Sigma_{\vec{\epsilon}}^{-1}.
\end{equation}
We now show that
\begin{equation}
\label{eq:Id-transform}
I-\Delta^{-1}(\C_\phi^{-1}+\Delta^{-1})^{-1}
=
(\Delta+\C_\phi)^{-1}\Delta.
\end{equation}
Let
\begin{equation}
A:=\C_\phi^{-1}+\Delta^{-1}.
\end{equation}
Since \(A\) is invertible, it is enough to prove that the two sides of Eq. \eqref{eq:Id-transform}
have the same product with \(A\) on the right. First,
\begin{equation}
\label{eq:Id-transform-left}
\Big(I-\Delta^{-1}A^{-1}\Big)A
=
A-\Delta^{-1}
=
\C_\phi^{-1}.
\end{equation}
Second,
\begin{align}
\label{eq:Id-transform-right}
(\Delta+\C_\phi)^{-1}\Delta\,A
&=
(\Delta+\C_\phi)^{-1}\Delta\big(\C_\phi^{-1}+\Delta^{-1}\big)
\\
&=
(\Delta+\C_\phi)^{-1}\big(\Delta\C_\phi^{-1}+I\big)
\\
&=
(\Delta+\C_\phi)^{-1}(\Delta+\C_\phi)\C_\phi^{-1}
\\
&=
\C_\phi^{-1}.
\end{align}
Therefore, Eq.~\eqref{eq:Id-transform} is established. Then, combining
Eq. \eqref{eq:GtSigmazinv_step1} and Eq. \eqref{eq:Id-transform} gives
\begin{equation}
\label{eq:GtSigmazinv_step2}
G^t\Sigma_{\mathbf{z}}(\phi)^{-1}
=
(\Delta+\C_\phi)^{-1}\Delta\,G^t\Sigma_{\vec{\epsilon}}^{-1}.
\end{equation}
Consequently,
\begin{equation}
H^t G^t \Sigma_{\mathbf{z}}(\phi)^{-1}
=
H^t(\Delta+\C_\phi)^{-1}\Delta\,G^t\Sigma_{\vec{\epsilon}}^{-1},
\end{equation}
and
\begin{equation}
H^t G^t \Sigma_{\mathbf{z}}(\phi)^{-1} G H
=
H^t(\Delta+\C_\phi)^{-1}\Delta\,(G^t\Sigma_{\vec{\epsilon}}^{-1}G)\,H
=
H^t(\Delta+\C_\phi)^{-1}H.
\end{equation}
Replacing these expressions into Eq.~\eqref{eq:gls_estimator} gives exactly Eq.~\eqref{eq:beta-woodbury}, which concludes the proof.
\end{proof}

\medskip
\noindent
From a numerical standpoint, the estimator in Eq. \eqref{eq:beta-woodbury} is advantageous when the dimension \(pm\) of \(\vec{\Theta}_m\) is small compared to the dimension \(nm\) of \(\vec{\mathbf{z}}\). The generalized least-squares expression in Proposition~\ref{prop4} requires the inversion of the marginal covariance matrix \(\Sigma_{\mathbf{z}}(\phi)\in\mathbb{R}^{nm\times nm}\), whereas the alternative formulation only involves the inversion of the matrix \(\Delta+\C_\phi\in\mathbb{R}^{pm\times pm}\). This reduction in matrix size leads to a significant decrease in computational cost when \(n\) is large.

\end{document}